\documentclass[usenatbib]{mnras}
\RequirePackage{rotating}
\usepackage{dcolumn}
\usepackage{bm}
\usepackage{lscape}
\usepackage{graphicx}
\usepackage{tabularx}
\usepackage{amsmath}
\usepackage{hyperref}
\usepackage{float}
\usepackage{siunitx}
\usepackage{braket}
\usepackage{color}
\usepackage{epsfig}
\usepackage{graphics}
\usepackage{longtable}
\usepackage{makecell}
\usepackage{array}
\usepackage[thinlines]{easytable}
\usepackage{multicol}

 at 13 truept
\def\fixv{\rlap{\phantom{$\strut^b$}}}

\begin{document}

\twocolumn

\title[Radio EBL]{The cosmic radio background from 150\,MHz--8.4\,GHz, and its division into AGN and star-forming galaxy flux}
\author[Tompkins et al.] 
{Scott A.\,Tompkins$^{1,2}$, 
Simon P.\,Driver$^{2}$,  
Aaron S.\,G.\,Robotham$^2$,
Rogier A.\,Windhorst$^1$,  \newauthor
Claudia del P.\, Lagos$^{2,3,4}$,
T. Vernstrom$^2$,
Andrew M.\,Hopkins$^{5}$\\
$^1$ School of Earth and Space Exploration, Arizona State University,
Tempe, AZ 85287-1404 \\
$^2$ International Centre for Radio Astronomy Research (ICRAR), University of Western Australia, Crawley, WA 6009, Australia \\
$^{3}$ARC Centre of Excellence for All Sky Astrophysics in 3 Dimensions (ASTRO 3D).\\
$^{4}$Cosmic Dawn Center (DAWN).\\
$^5$ Australian Astronomical Optics, Macquarie University, 105 Delhi Rd, North Ryde, NSW 2113, Australia}

\pubyear{2023} \volume{000}
\pagerange{\pageref{firstpage}--\pageref{lastpage}}

\maketitle
\label{firstpage}

\vspace{-2.0cm}

\begin{abstract}
We present a revised measurement of the extra-galactic background light (EBL) at radio frequencies based on a near complete compendium of radio source counts. We present the radio-EBL at 150 MHz, 325 MHz, 610 MHz, 1.4 GHz, 3 GHz, 5 GHz, and 8.4 GHz. In all cases the contribution to the radio-EBL, per decade of flux, exhibits a two-humped distribution well matched to the AGN and star-forming galaxy (SFG) populations, and with each population contributing roughly equal energy. Only at 3 GHz are the source count contributions to the EBL fully convergent, and hence we report empirical lower limits to the radio-EBL in the remaining bands. Adopting predictions from the SHARK semi-analytic model for the form of the SFG population, we can fit the fainter source counts providing measurements of the total contribution to the radio-EBL for the SFG and the AGN populations separately. This constitutes an empirically constrained model-dependent measurement for the SFG contribution, but a fully empirical measurement of the AGN contribution. Using the {\sc ProSpect} spectral energy distribution code we can model the UV-optical-infrared-mm-radio SFG EBL at all frequencies from the cosmic star-formation history and the adoption of a Chabrier initial mass function. However, significant discrepancy remains ($5\times$) between our source-count estimates of the radio-EBL and the direct measurements reported from the ARCADE-2 experiment. We can rule out a significant missing discrete source radio population and suggest that the cause of the high ARCADE-2 radio-EBL values may need to be sought either in the foreground subtraction or as a yet unknown diffuse component in the radio sky.
\end{abstract}

\begin{keywords}
surveys,
radio continuum: galaxies,
galaxies:active,
cosmology: cosmic background radiation,
cosmological parameters, catalogues
\end{keywords}

\section{Introduction}
In recent years there has been a resurgence of interest in measurements and studies of the Extra-galactic Background Light or EBL, (e.g., \citealt{Gervasi_2008}; \citealt{Vernstrom_2011}; \citealt{Driver_2016}; \citealt{Abdallah_2018};  \citealt{Lauer_2022}; Saldana-Lopez et al. 2022, submitted). The EBL is the term used to describe all radiation incident on the Earth of extra-galactic origin from a steradian of sky, i.e., it should exclude any sky-glow, Zodiacal and Diffuse Galactic Light components, as well as any light from the Milky-Way group. In terms of the origin of the EBL it can be divided into radiation arising from two eras (epochs): the Cosmic Microwave Background which represents relic radiation from the hot early Universe at the time of recombination and redshifted to the present day; and the remainder which is all radiation produced from all eras since recombination. The latter predominantly arises from star-formation, accretion onto super-massive black holes (i.e., Active Galactic Nuclei; AGN), and dust reprocessing which typically transfers ultraviolet and optical radiation to the mid and far infrared as thermal emission. 

For convenience the EBL is also often broken into distinct wavelength regimes that cover the cosmic $\gamma$-ray (CGB), X-ray (CXB), ultraviolet (CUB), optical (COB), infrared (CIB), the CMB, and radio (CRB) backgrounds. Of these backgrounds the CMB is the strongest, in terms of its integrated energy density, and approximately 5$\times$ the sum of the other backgrounds combined \citep{Hill_2018}. Of these other backgrounds the COB and CIB are roughly equal \citep{Driver_2016}, and together comprise most of the non-CMB energy contribution.

The recent resurgence of interest in the EBL, is in part due to technological breakthroughs that have allowed the measurement of the various backgrounds to much lower flux limits within each waveband. This has allowed measurements to evolve from upper or lower limits to credible measurements. This is possible because in almost all bands we are now able to resolve the peak contribution to the EBL by constructing galaxy or source counts that show the number-density of discrete sources as a function of flux (see recent summary by \citealt{Hill_2018}). Overall, our measurements of the EBL now extend from $\gamma$-rays (VHE; \citep{Biteau_2015, Abeysekara_2019, Abdollahi_2018} through to X-ray \citep{Giacconi_1962, Gilli_2007}, the optical and infrared \citep{Finke_2010, Driver_2016}, and ultimately into the radio \citep{Murphy_2018, de_Zotti_2009, Fixsen_2011, Vernstrom_2011}. 

However, the resurgence also stems from the growing appreciation that many of the physical phenomena we wish to study (e.g., star-formation and AGN), result in photon production that manifests across many wavelengths. For example, star-formation results in energy production from X-ray binaries \citep{Natarajan_2000, Khaire_2019}, supernova explosions, thermal radiation, synchrotron and free-free emission \citep{Condon_1984,Condon_1992,Dale_2014}. As a consequence the latest generation of Spectral Energy Distribution fitting models (e.g., CIGALE, \citeauthor{Boquien_2019}, \citeyear{Boquien_2019}; ProSpect, \citeauthor{Robotham_2020} \citeyear{Robotham_2020}), as well as the latest generation of semi-analytic simulations \citep{Lagos_2018, Inoue_2013, Baes_2019} now extend from the X-ray to radio domains.

This insight follows from our understanding that star-formation inferred in the optical/near-IR can now be used to accurately predict the radio continuum fluxes \citep{van_der_Kruit_1971, Davies_2017} and the inverse. For example, estimates of the cosmic star-formation history \citep{Madau_2014} can now be obtained from Very High Energy studies of Blazars and the interaction of their flux with the CIB \citep{Abdollahi_2018}, or from CRB constraints \citep{oldMatthews_2021, Nitu_2021} alone. Similarly, AGN are known to produce radiation at all wavelengths albeit in a stochastic manner, in which none, one, two or all three wavelength regimes (X-ray, optical/IR and radio) might be active or not at any one time. Hence to fully understand the AGN life-cycle and their role within galaxy formation also necessitates panchromatic astronomy from X-ray to radio wavelengths.

The long-term potential of EBL studies is that if we can truly understand the history of star-formation, the evolution of super-massive black holes, and the role of dust processing, then we should be able to explain the entire cosmic photon flux incident on the Earth. This would be a remarkable achievement. Even more promising perhaps, is the inverse, whereby accurate measurements of the EBL and its subdivision into redshift slices (a.k.a., the Cosmic Spectral Energy distributions) might allow us to reconstruct the entire cosmic star-formation history and its dependencies on model ingredients such as the initial mass function and metallicity evolution, for example, see early attempts to do this by \citet{Koushan_2021} or Saldana-Lopez et al. (2022, submitted).

However, at present there is tension in the EBL measurements between direct and indirect methods. See for example the review by \citet{Driver_2021}. This manifests most in the COB, CIB and CRB in which direct and indirect measurements disagree by factors of $\times 3-10$. Direct measurements typically necessitate the measurement of the "above Earth sky" followed by subtraction of foregrounds such as the Zodiacal Light and Diffuse Galactic Light for the COB and CIB, and subtraction of radio radiation emerging from the Galactic halo and surrounding high-velocity clouds and removal of the CMB for the CRB estimate \citep{Fixsen_2011}. Indirect estimates include the measurement and modelling of galaxy number-counts (COB, CIB), or source counts (CRB), which are only sensitive to discrete sources of radiation (i.e., galaxies and AGN) but benefit by not being plagued by overwhelming foregrounds (COB, CIB) or backgrounds (CRB). 

In the optical, the difficulty in the direct measurements likely stems from the subtraction of the Zodiacal Light and Diffuse Galactic Light which are typically 10-100$\times$ the level of the EBL and dependent on direction and time of observation within the year. There is also a growing appreciation of the impact of sky-glow  \citep{Caddy_2022} on direct background estimates from facilities such as the Hubble Space Telescope (\textit{HST}). In the COB and CIB the long standing factor of $\sim 10$ discrepancy (see for example \citealt{Bernstein_2002}) has possibly now been resolved by indirect constraints from Very High Energy constraints (e.g.,  \citealt{Aharonian_2006};  \citealt{Ahnen_2016}; \citealt{HESS_2013}; \citealt{Abdollahi_2018})  appearing to corroborate the low EBL results from integrated source counts. This also follows from a deeper understanding of the Zodiacal Light from recently revised \textit{HST} Sky-Surface brightness measurements \citep{Windhorst_2022}. However some level of tension between the direct and indirect COB does still remain as recent results from the New Horizons instrument, from distances beyond where Zodiacal Light should be significant, still show a factor of $\times$2 inconsistency with the indirect count and VHE constraints, \citep{Lauer_2021, Lauer_2022}.

In the radio there also appears to be an immutable discrepancy of a factor of 2-5$\times$ between the (high) direct measurements from instruments such as the Absolute Radiometer for Cosmology, Astrophysics and Diffuse Emission (ARCADE-2; \citealt{Fixsen_2011}), and recent source count studies (see for example \citealt{Vernstrom_2011}). In the radio the critical subtraction for these direct measurements is the Raleigh-Jeans tail of the CMB which still outshines the radio source count signal at high frequencies. However, a compounding problem is that most source counts are not quite deep enough to become convergent. Hence, source count estimates of the CRB require sophisticated modeling of the SFG population (see for example  \citealt{Wilman_2008};  \citealt{Hopkins_2003}; \citealt{Seymour_2008};  \citealt{Seymour_2004} ; \citealt{Mancuso_2017}).

In this paper, which forms part of our broader work on the overall EBL, we revisit the current constraints on the CRB from source count data across a broad frequency range, and building on earlier studies and compendiums by \citet{Windhorst_2003}, \citet{Vernstrom_2011}, and \citet{de_Zotti_2009}. This is motivated on the basis of extending our earlier optical-IR motivated models into the radio \citep{Andrews_2018, Koushan_2021}, and also in preparation for upcoming surveys scheduled to take place on a myriad of new radio facilities (i.e., ASKAP, MeerKAT, LOFAR, MWA etc) in anticipation of the upcoming Square Kilometer Array \citep{Bonaldi_2019}. These facilities, and the SKA especially, should have the capacity to finally extend source counts to sufficiently low flux limits for the CRB to be convergent at all frequencies without recourse to models or extrapolations. Hence, this paper is intended to act as a useful reference of previous works, as well as a launchpad for future studies using imminent new technologies.

In the later stages of the paper we use the recent SHARK semi-analytic simulation \citep{Lagos_2018} to model the SFG population to allow us to extrapolate our source counts in a meaningful manner by including the star-formation population that lies below our detection thresholds. This also opens the door for us to constrain the CRB using the model predictions to separate out the CRB contribution from the AGN and SFG populations separately.

In Section 2 we describe the compendium of radio data that we have assembled across 7 frequency bands: 150\,MHz, 325\,MHz, 610\,MHz, 1.4\,GHz, 3\,GHz, 5\,GHz and 8.4\,GHz and we describe our method for implementing spectral index corrections when necessary and show our source-counts, source counts normalized to Euclidean ($N/S^{-2.5}$), and the more useful rendition that shows the contribution of each flux interval to the total energy density ($N/S^{-2}$). In Section 3 we describe our attempts to extract the CRB for the AGN and SFG using simple spline fitting with and without extrapolation. In Section 4 we describe a more robust method for extracting the SFG and AGN CRB contributions by fitting of the recent SHARK semi-analytic model \citep{Lagos_2018}. Finally, in Section 5 we show how our ProSpect EBL model extends into the radio and how good the cosmic star-formation history is at matching the CRB for the SFG population.

We present our conclusions in Section 6.
Throughout we adopt a standard cosmology of $H_o=70$ km/s/Mpc, $\Omega_{\rm M}=0.3$ and $\Omega_{\Lambda}=0.7$.

\section{Radio surveys, catalogues and data}
Here we aim to provide a complete compendium of radio source count data covering 7 frequencies: 150\,MHz, 325\,MHz, 610\,MHz, 1.4\,GHz, 3\,GHz, 5\,GHz, 8.4\,GHz. Note that this compendium does not include all the recent 887.5 and 943.5 MHz wavelength data as this is only just emerging from ASKAP via surveys such as Evolutionary Map of the Universe (EMU; see for example \citealt{Norris_2021}; \citealt{Gurkan_2022}). In total, our final compendium is comprised of 74 distinct radio surveys spread across 16 different frequencies and corrected to one of the 7 frequencies listed above. This compendium is designed to cover the largest flux density range possible in each frequency band as a baseline reference for future survey programs. The data hence includes large all-sky surveys such as VLASS \citep{Gordon_2021}, NVSS \citep{Condon_1998}, and LoTSS \citep{Hardcastle_2021} which provide catalogues often numbering into the millions of objects, and excellent statistics for bright source counts. These are complemented by deep small area surveys which reach to very faint flux levels extending down and slightly below ~1\,mJy. \\

The data assembled in this paper were taken with many different instruments and at many different frequencies over the last 5 decades. Early data was typically taken by single-dish radio telescopes, such as the National Radio Astronomy Organization (NRAO) 300-ft telescope, which provides some of the oldest source counts originally assembled as part of the \citet{Windhorst_2003} compendium. This data provides the brightest source counts at 1.4\,GHz, and which were used for the first demonstration of the expected Euclidean behavior of the density of bright source counts in the local universe. Multi-dish synthesis telescopes such the Westerbork Radio Synthesis Telescope (WSRT), the Australian Compact Telescope Array (ATCA), and the Very Large Array (VLA) provide much of the bright source counts in the modern era, achieving finer resolution than is possible with single-dish telescopes and extending to higher frequencies. Similar arrays such as the Giant Meterwave Radio Telescope (GMRT) provide a similar benefit at lower frequencies. Additional counts are also provided by the latest generation of radio facilities that include LOFAR, MWA, ASKAP and MeerKAT (with the latter three representing the SKA precursors).

Our data also builds on previous compendiums of radio source count data such as that provided by \citet{Windhorst_2003}  and \citet{Vernstrom_2011}. In particular the \citet{Vernstrom_2011} study  used the data available at the time to make a state-of-the art multi-frequency estimate of the discrete radio EBL covering six frequencies. These compendiums and the references therein provided the starting point for gathering the source count data described in the current paper. 

In assembling our final data compendium, we adjusted all of the data to the same format, essentially frequency and source density per steradian, along with upper and lower limits with additional information such as central frequency, area and survey labels. Hence we have constructed a single homogenized table of source counts covering all contributing surveys and frequencies and which we provide as a Machine Readable Table, a sample of which is shown in the Appendix as Table \ref{tab:A.4}.

In Table\, \ref{tab:inputdata}, we provide a summary of the data sets used in this paper, and include in the Appendix a breakdown of the compendium from \citet{Windhorst_2003} as Table \ref{tab:A.1}. In these tables we provide the original frequency, reference, instrument, and area surveyed in steradians, for all data sets used. Where possible, we provide the field or survey name and appropriate field center. Finally, we provide the Full Width at Half Maximum ({\sc fwhm}) of the radio survey instrument's beam in arcseconds. When given in the paper, the stated value for the beam is used. If the beam was not specified, it is calculated by using {\sc fwhm} $= 1.22\times \frac{\lambda}{D}$ where $\lambda$ is the central wavelength of the receiver bandwidth used and $D$ is the telescope diameter or maximum baseline, both in meters, where {\sc fwhm} in radians is converted to arcseconds.

\begin{table*}
\label{tab:data}
\caption{Summary of the various data sets used in this paper.
\label{tab:inputdata}}
\begin{tabular}{p{1.0cm}p{3.0cm}p{1.5cm}p{2.0cm}p{3.0cm}p{1.5cm}p{1.5cm}p{2.0cm}} \\ \hline
Frequency  & Reference & Instrument & Field/Survey  & Field Location & Area & Beam FWHM & Confusion Source Density/ \\ 
(MHz) & & & Name & & (Ster) & (Arcsec)  & Faintest Source Bin (mJy) \\ \hline
154 & \citet{Franzen_2016} & MWA & MWA EoR0 & $\alpha = 00^h 00^m 00^s,\delta = -27^{\circ} 00^{'} 00^{''}$ & 0.1736 & 138.6 & $1.14\times 10^{-2}$ \newline  33.8 \newline\\ 
150 & \citet{Mandal_2021} & LOFAR & Lockman; Hole Boötes; ELAIS-N1 & Multiple Fields & $2.31\times 10^{-2}$ & 6 & 1.35 \newline  0.22  \\ 
151 & \citet{McGilchrist_1990} & CLFST & Unnamed Fields & 2 Field Centers & 0.144 & 70 & $1.32\times 10^{-3}$ \newline  105.9 \newline\\ 
151.5 & \citet{Hardcastle_2021} & LOFAR & LoTSS-DR2, Boötes, ELAIS-N1, Lockman & Multiple & 0.6 (LoTSS-DR2) $2.78\times 10^{-3}$ (Summed total of deep fields) ,  & 6 & 0.106 \newline 0.01 \\ 
150 & \citet{Williams_2016} & LOFAR & Boötes \& Multiple Pointing Centers & Boötes - $\alpha = 14^h 32^m 00^s,\delta = +34^{\circ} 30^{'} 0^{''}$ & $5.79\times 10^{-3}$ & 4.19  & $3.07\times 10^{-4}$ \newline  0.71\\
200 & \citet{Hurley-Walker_2016} & MWA & GLEAM & All-Southern Sky Excluding Galactic Plane and Magellanic Clouds & 7.5640 & 120 & $1.02\times 10^{-4}$ \newline  25\\ 
325 & \citet{Mazumder_2020} & GMRT & Lockman Hole & $\alpha = 10^h 48^m 00^s,\delta = 58^{\circ} 08^{'} 00^{''}$ & $1.83\times 10^{-3}$ & 9 & $1.42\times 10^{-3}$ \newline  0.4 \newline\\ 
327 & \citet{Oort_1988} & WSRT & Lynx &  $\alpha = 08^h 39^m 52.5^s,\delta = +43^{\circ} 52^{'} 44^{''}$  & $1.54\times 10^{-2}$ & 64.7 & $1.67\times 10^{-2}$ \newline  6.71 \newline \\ 
325 & \citet{Riseley_2016} & GMRT & Super-Class; Abell & Multiple Pointing Centers & $1.98\times 10^{-3}$ & 13 & $5.29\times 10^{-3}$ \newline 0.242\newline\\
325 & \citet{Sirothia_2009} & GMRT & ELAIS-N1 & $\alpha = 16^h 10^m 00^s,\delta = 54^{\circ} 36^{'} 00^{''}$ & $7.62\times 10^{-5}$ & 8.31 & $1.35\times 10^{-3}$ \newline 0.367\newline\\ 
610 & \citet{Bondi_2006} & GMRT & VVDS-VLA & $\alpha = 02^h 26^m 00^s,\delta = -04^{\circ} 30^{'} 00^{''}$ & $3.05\times 10^{-4}$ & 6 & $4.20\times 10^{-4}$ \newline 0.37\newline\\ 
610 & \citet{Garn_2008} & GMRT & ELAIS-N1 & $\alpha = 16^h 11^m 00^s,\delta = 55^{\circ} 00^{'} 00^{''}$ & $3.05\times 10^{-4}$ & 5.477 & $3.94\times 10^{-5}$ \newline 0.338\newline\\ 
887.5 & \citet{Hale_2021} & ASKAP & RACS & All-Sky South of  $\delta = 41^{\circ} 00^{'} 00^{''}$ & 8.535 & 25 & $1.46\times 10^{-3}$ \newline 1.40\newline\\ 
610 & \citet{Ibar_2009} & GMRT & Lockman Hole & Multiple Pointing Centers & $2.99\times 10^{-4}$ & 4.25 & $2.34\times 10^{-3}$ \newline 0.056\newline\\ 
610 & \citet{Moss_2007} & GMRT & 1H XMM-Newton/Chandra & $\alpha = 1^h 11^m 00^s,\delta = -4^{\circ} 00^{'} 00^{''}$ & $2.72\times 10^{-4}$ & 6.69 & $3.53\times 10^{-4}$ \newline 0.49\newline\\ 
610 & \citet{Ocran_2020}& GMRT & ELAIS-N1 & $\alpha = 16^h 10^m 30^s,\delta = 54^{\circ} 35^{'} 00^{''}$ & $5.66\times 10^{-4}$ & 6 & $2.97\times 10^{-3}$ \newline 0.078\newline\\ 
1400 & \citet{Biggs_2006} & VLA & HDFN, ELAIS-N1 & Multiple Pointing Centres & $8.12\times 10^{-5}$ & 1.51 & $2.30\times 10^{-4}$ \newline 0.402\newline\\
1400 & \citet{Condon_1998} & VLA & NVSS & All-Sky North of  $\delta = -40^{\circ} 00^{'} 00^{''}$ & 11.34 & 45 & $2.06\times 10^{-3}$ \newline 3.16\newline\\ 
1400 & \citet{Fomalont_2006} & VLA & SSA13 & $\alpha = 13^h 12^m 2,\delta = +42^{\circ}38^{'} 0^{''}$ & $9.78\times 10^{-5}$ & 1.54 & $2.92\times 10^{-4}$ \newline 0.034\newline \\ 
1400 & \citet{Hopkins_2003} & ATCA & Phoenix Deep Survey-Deep & $\alpha = 01^h 14^m 12^s,\delta = +45^{\circ} 44^{'}8^{''}$ & $9.21\times 10^{-5}$ & 8.48 & $4.69\times 10^{-3}$ \newline 0.057\newline\\ 

\end{tabular}
\end{table*}

\setcounter{table}{0}

\begin{table*}
\caption{(cont'd) Summary of the various data sets used in this paper.}
\begin{tabular}{p{1.0cm}p{3.0cm}p{1.5cm}p{2.0cm}p{3.0cm}p{1.5cm}p{1.5cm}p{2.0cm}} \\ \hline
Frequency  & Reference & Instrument & Field/Survey  & Field Location & Area & Beam FWHM & Confusion Source Density \\ 
(MHz) & & & Name & & (Ster) & (Arcsec)  & Faintest Source Bin (mJy) \\ \hline

1400 & \citet{Hopkins_2003} & ATCA & Phoenix Deep Survey-Primary & $\alpha = 01^h 14^m 12^s,\delta = +45^{\circ} 44^{'}8^{''}$ & $1.37\times 10^{-3}$ & 8.48 & $2.34\times 10^{-3}$ \newline 0.1\newline \\ 
1400 & \citet{Huynh_2005} & ATCA & HDFS & $\alpha = 22^h 33^m 25^s,\delta = - 60^{\circ} 38^{'} 09^{''}$ & $1.06\times 10^{-4}$ & 6.6 & $2.65\times 10^{-3}$ \newline 0.059\newline\\
1400 & \citet{Matthews_2021} & MeerKAT & DEEP2 & $\alpha = 04^h 13^m 26^s, \delta=-80^{\circ} 0^{''} 0^{'}$ & $3.17\times 10^{-4}$ & 7.6 & $2.56\times 10^{-2}$ \newline 0.0126\newline\\ 
1400 & \citet{Prandoni_2000} & ATCA & ATESO & Multiple Pointing Centres & $7.92\times 10^{-3}$ & 7.6 & $3.33\times 10^{-4}$ \newline 0.83\newline\\ 
1400 & \citet{Seymour_2008} & VLA & $13^h$ Chandra Deep Field & $\alpha = 13^h 34^m 37^s,\delta = +37^{\circ} 54^{'} 44^{''}$ & $5.97\times 10^{-5}$ & 3.3 \\ 
1400 & \citet{Windhorst_2003} & Multiple & N/A & Large Data Compendium & See Appendix for Windhorst compendium information &  \\ 
2100 & \citet{Butler_2017}& ATCA & XXL-S & $\alpha = 23^h 30^m 00^s,\delta = -55^{\circ} 0^{'} 0^{''}$ & $7.62\times 10^{-3}$ & 4.76 & $1.12\times 10^{-4}$ \newline 0.338 \newline\\ 
3000 & \citet{Gordon_2021} & VLA & VLASS & Entire Sky North of $\delta = -40^{\circ} 00^{'} 00^{''}$ & 10.75 & 2.5 & $5.30\times 10^{-6}$ \newline 2.50 \newline\\ 
3000 & \citet{Smolcic_2017} & VLA & COSMOS & $\alpha = 10^h 00^m 28.6^s,\delta = - 02^{\circ} 12^{'} 21^{''}$ & $6.09\times 10^{-4}$ & 0.75 & $1.41\times 10^{-4}$ \newline 0.011 \newline\\ 
3000 & \citet{Van_der_Vlugt_2021} & VLA & COSMOS & $\alpha = 10^h 00^m 28.6^s,\delta = - 02^{\circ} 12^{'} 21^{''}$ & $1.52\times 10^{-5}$ & 1.97 & $7.02\times 10^{-3}$ \newline $3.72\times 10^{-3}$ \newline\\ 
3000 & \citet{Vernstrom_2016} & VLA & Lockman Hole & $\alpha = 10^h 46^m 00^s,\delta = +59^{\circ} 0^{'} 0^{''}$ & $1.60\times 10^{-5}$ & 8 & $1.36\times 10^{-2}$ \newline 0.013 \newline\\ 
5500 & \citet{Huynh_2015} & ATCA & Chandra Deep Field South & $\alpha = 3^h 32^m 28.0^s,\delta = - 27^{\circ} 48^{'} 30^{''}$ & $7.62\times 10^{-5}$ & 3.16 & $2.60 \times 10^{-4}$ \newline 0.051 \newline\\ 
5000 & \citet{Prandoni_2006} & ATCA & ASTEP & Large Survey Near SGP at $\alpha = 22h,\delta = - 40^{\circ}$ \& & $3.05\times 10^{-4}$ & 10.1 & $3.10\times 10^{-4}$ \newline 0.44 \newline\\ 
4850 & \citet{Windhorst_2003} & Multiple & N/A & Large Data Compendium & See Appendix For Windhorst Compendium Information &  \\ 
8500 & \citet{Henkel_2005} & VLA & HDFN & $\alpha = 12^h 36^m 49.4^s,\delta = - 62^{\circ} 12^{'} 58^{''}$ & $4.57\times 10^{-4}$ & 2.28 & $1.92\times 10^{-5}$ \newline 0.21 \newline\\ 
9000 & \citet{Huynh_2020} & ATCA & Chandra Deep Field South & $\alpha = 3^h 32^m 28.0^s,\delta = - 27^{\circ} 48^{'} 30^{''}$ & $8.41\times 10^{-5}$ & 2.28 & $2.52\times 10^{-4}$ \newline 0.109 \newline\\ 
8400 & \citet{Windhorst_2003} & Multiple & N/A & Large Data Compendium & See Appendix For Windhorst Compendium Information &  \\ \hline
\end{tabular}
\end{table*}

\clearpage

On Figure \ref{fig:fig2} we show our compendium of counts for 3\,GHz and the equivalent figure for all other frequencies are shown in the Appendix as Figure\,\ref{fig:A1}. We show in the top panels the source counts, in the middle panel the source counts divided by the Euclidean slope ($S^{-2.5}$), and in the lower panel the source counts divided by $S^{-2}$ and which best shows the contribution of each flux interval to the overall EBL in that frequency. For the 3\,GHz data shown on Figure\,\ref{fig:fig2} we can see that in the lower panel we have a two humped distribution, reflecting the contribution from AGN (right hump) and SFG (left hump) with each providing a roughly equal contribution to the total EBL (the full integral under the curve.)

\subsection{Spectral index corrections}
\label{Section_2.1}
In order to compare radio data taken at slightly different frequencies, corrections must be applied. For example, the 1.4\,GHz source counts have been traditionally obtained at frequencies ranging from 1.4--1.5\,GHz. Hence corrections need to be applied to bring the central frequency of each survey onto a common scale. Corrections are achieved using an adopted spectral index, hereafter referred to as $\alpha$ where the flux, $S_{\nu} \propto \nu^{-\alpha}$. In most previous work, a constant or median value of $\alpha$ is adopted for simple visual comparisons. This is often assumed to have a median value of $\alpha \sim 0.7$. However, the spectral index is known to vary as a function of radio flux. This is because the radio source counts are dominated by different populations at different fluxes which do not necessarily have the same nature, age, or redshift. Hence, assuming a constant value for $\alpha$ over all frequency and flux ranges is not necessarily correct. For the frequency corrections used in this paper, we allow $\alpha$ to vary as a function of flux as detailed in \citet{Windhorst_2003}. We then use the \citet{Windhorst_2003} relationships between the median $\alpha$, frequency and radio flux to perform our corrections. This adjustment should bring all surveys within our 7 frequency bands into frequency alignment. 

\begin{figure}
\vspace*{0.000cm}
\includegraphics[width=\columnwidth]{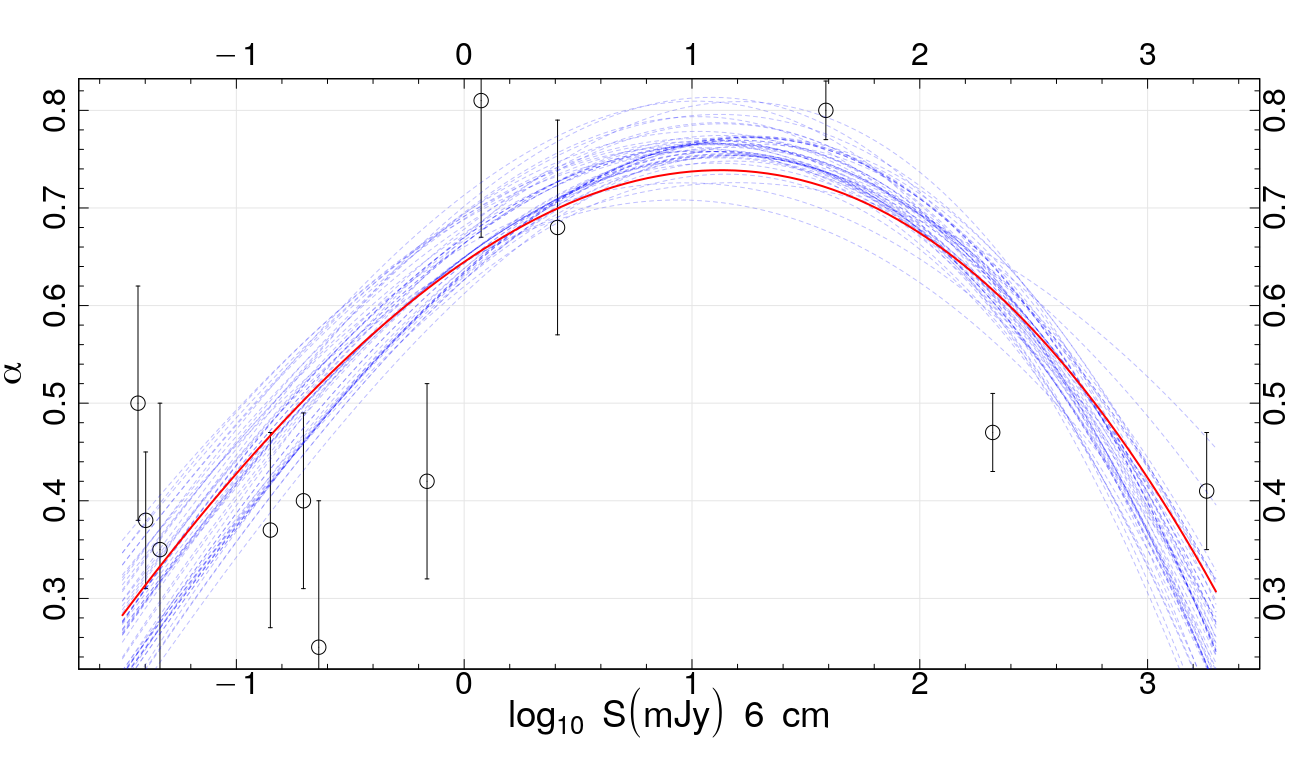}
\includegraphics[width=\columnwidth]{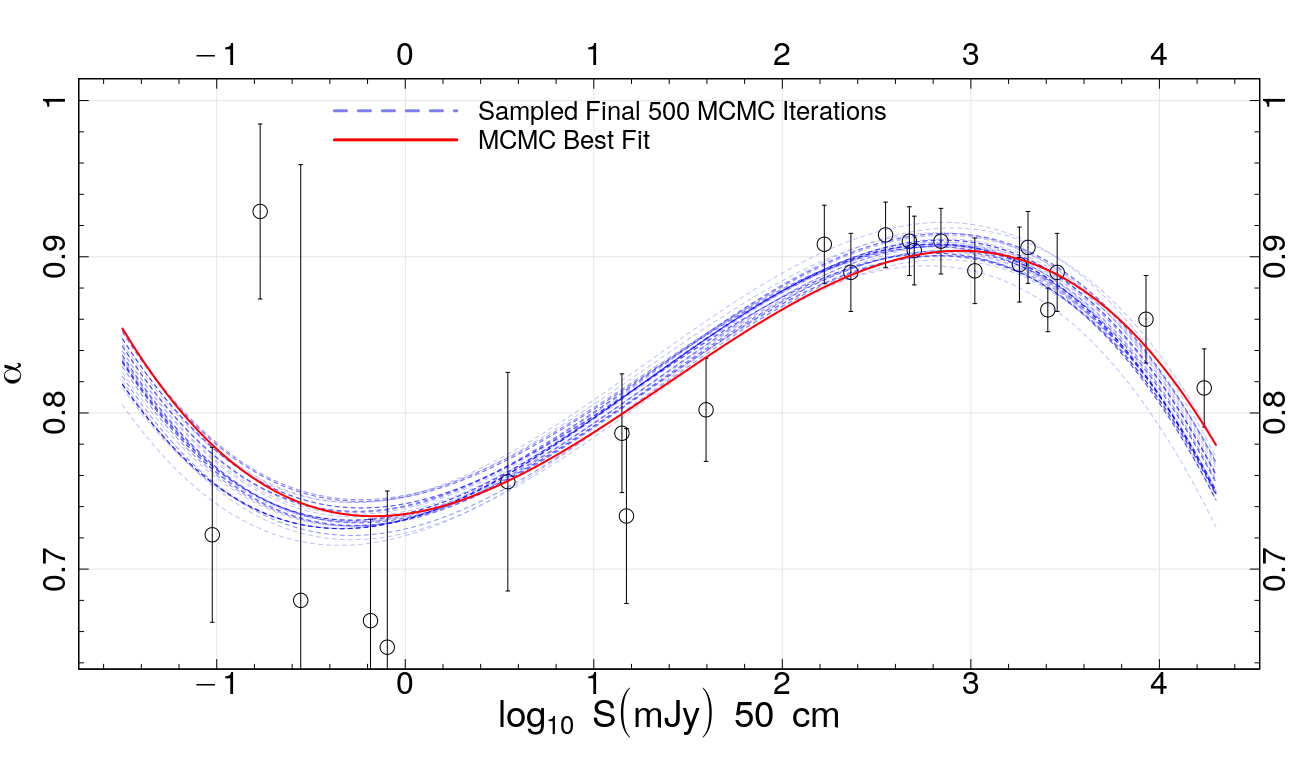}

\vspace*{0.00cm}
\caption{(upper Fig. 1a) The median spectral index $\alpha$ versus flux relationship at a wavelength of 6cm or frequency of ~5\,GHz.  (Lower Fig. 1b) The median $\alpha$ versus flux relationship at a wavelength of 50 cm or frequency of ~610\,MHz. A sample from the final 50 iterations of the MCMC analysis are shown in blue and the best-fit is shown in red.}
\label{fig:fig1}
\end{figure}

\begin{figure*}
\vspace*{0.000cm}
\includegraphics[width=16cm, height=19.5cm]{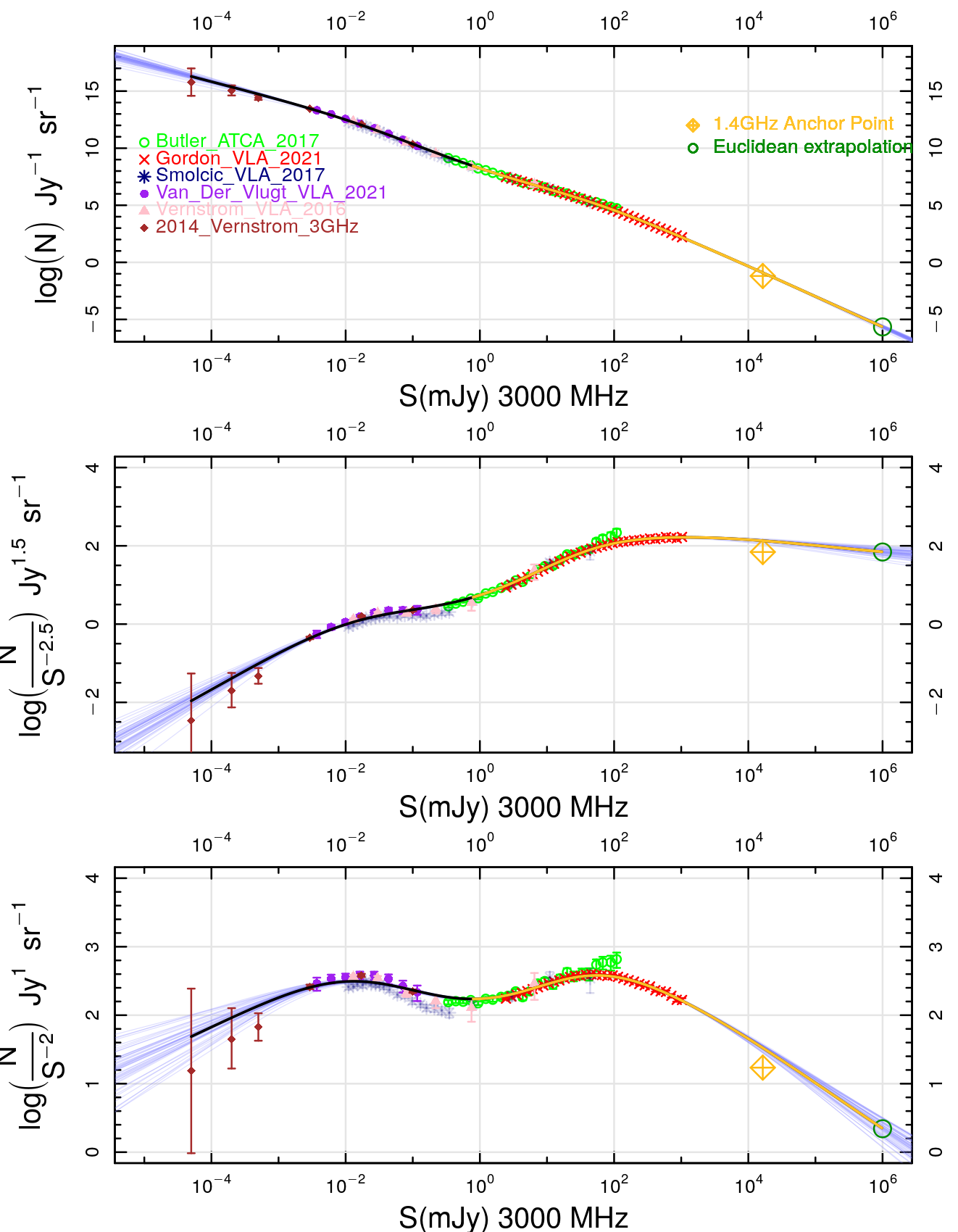}

\caption{(upper) A collection of differential 3.0\,GHz radio source counts. The error bars on most points are too small to display in this format, however, they are still present on all points. (centre) A collection of normalized 3.0\,GHz differential radio source counts spanning $\sim$7 orders of magnitude.  (lower) The same collection of differential 3.0\,GHz radio source counts showing the radio EBL. The counts are normalized to $\log_{10}(N/S_v^{-2.0}$) to highlight contributions to the radio EBL. The right peak centered near $\sim10^{2.5}$\,mJy displays the contribution of primarily AGN and the left peak beginning to form near $\sim10^{-2}$\,mJy. The dark black and yellow line is the best spline fit to the data while the light blue lines are spline fits which were allowed to vary within the error bars of each data point and the error floor added to each point. The yellow portion of the best fit line outlines the first peak of the radio EBL contribution dominated by AGN. The convergence at both ends allows for a convergent measurement of the discrete radio EBL. The faintest three points from 
\citet{Vernstrom_2014} 
provide this convergence despite being a statistical analysis of the noise, which is reflected by the large error bars and uncertainty in the converging slope. Points with transparency are shown for completeness but are left out of the spline fit for a variety of reasons detailed in Sections \ref{Section_2.2} and \ref{Section_2.3}. The same figures for the other frequencies are shown in the appendix.
\label{fig:fig2}}
\end{figure*}

Figure\,\ref{fig:fig1} shows the median $\alpha$ at high (upper panel) and low (lower panel) frequencies as a function of radio flux and we fit these data with cubic polynomials to determine the optimum value of $\alpha$ for each flux point. The spectral index $\alpha$ was measured  between 0.41 or 1.4 GHz and 4.86 or 5.0 GHz. Many of the median $\alpha$ values summarized in \citet{Windhorst_2003} were originally measured by \citet{Kapahi_1986}, who used sources identified at 0.41 and 5.0 GHz to derive a median spectral index between the two frequencies. Care was taken to only use surveys that had the FWHM of the survey frequency with the highest resolution convolved to become equal to the FWHM of the survey frequency with the lower resolution, so as to not bias the spectral index measurements and their medians that were made as a function of flux density. We also made sure to use only sources above the 5-$\sigma$ detection limits at {\it both} survey frequencies in order not to bias the median spectral index derived as a function of radio flux. The flux scales of the actual surveys were then converted to two fiducial frequencies, i.e., either 610 MHz or 4.86 GHz, whichever was closest to the observed frequency of each survey. This was done using the relationships derived in \citet{Windhorst_2003}, which we have refined in Figure\,\ref{fig:fig1}. Of course, this process requires the actual relationship to be known before it can be used to convert the flux scales at adjacent frequencies to the fiducial frequency, and so we iterated the best fits in Figure\,\ref{fig:fig1} until convergence was reached, which happened quickly within 1-2 iterations. For surveys not done at 4.86 GHz or 610 MHz, the relationship closer in frequency to a data set was used to perform these frequency corrections. Figure\,\ref{fig:fig1} highlights that using a constant spectral index to bring radio surveys at slightly different frequencies onto the same scale can be inaccurate.

A cubic polynomial is fit to each $\alpha$ vs $\log_{10}$ flux relationship in Figure \ref{fig:fig1}. The fits are generated via Monte-Carlo Markov-Chain analysis using the \textsc{LaplacesDemon} software package. 


\vspace*{0.00cm}

\subsection{Problematic data sets}
\label{Section_2.2}
Due to the significant amount of data used --- extending back to the 1970s and using a range of instruments, frequency corrections (where applicable), areas, resolutions, and survey regions --- inconsistencies are to be expected. During our review of the data only two entire data sets were felt to be sufficiently anomalous to be omitted from our final analysis but are included in the compendium for completeness. At 150\,MHz, the data of \citet{McGilchrist_1990} is not considered in our fitting, since the newer data set from the LOFAR Two-Meter Sky Survey of \citet{Mandal_2021} covers the same range over a larger area, while being fully consistent with other available data. At 3\,GHz, the data of \citet{Smolcic_2017} is also shown, but again not included in the fitting. This relatively recent VLA survey was conducted in the COSMOS field, and the inconsistency between their data and those of \citet{Van_der_Vlugt_2021} is addressed in the latter paper, where the source counts of \citet{Van_der_Vlugt_2021} are consistently a factor of ~1.4 above those of \citet{Smolcic_2017}. This offset is discussed by \citet{Van_der_Vlugt_2021} and partially ascribed to the impact of resolution bias as the Van der Vlugt team used a higher resolution ($0.2^{''}$ versus $0.75^{''}$) and hence less prone to missing sources due to resolution bias. The other contribution to the offset explored by \citet{Van_der_Vlugt_2021} could be due in part to cosmic variance, though this is not claimed to be the dominant effect. In this case the \citet{Smolcic_2017} data covered a larger area than \citet{Van_der_Vlugt_2021}, the latter of which surveyed an area completely contained within the former. For a more complete discussion and analysis of the two data sets see \citet{Van_der_Vlugt_2021}. Given the incompatibility of these data sets we elected to adopt the \citet{Van_der_Vlugt_2021} data purely on the basis of its greater consistency with two additional independent data sets (see Figure \ref{fig:fig2}).

\subsection{Problematic data points}
\label{Section_2.3}
For many of the data sets,  assembled data points are often included that suggest incompleteness at the faint end --- i.e., the source counts flatten or turn down compared to a best fit extrapolation of the brighter survey points that are very likely complete. The faintest flux bins in each survey have typically 50--100 sources per bin, and therefore RMS errors of the order of 14--10\%, respectively in each of the faintest bins. As a working definition, therefore, we would consider faint object counts to have become visibly incomplete if the faintest bins in each survey are higher or lower than the (power-law) extrapolation from brighter bins by twice that amount, or 28--20\%, respectively. Such faintest flux bins have subsequently been flagged as potentially incomplete in that survey, and are therefore not used in our fits of the sources counts.
For each dataset we need to determine the limits over which the data is credible. In most cases, the majority of a data set is used, with some very bright or very faint data points flagged as incomplete or of insufficient accuracy. At low flux limits this is due to the survey sensitivity and at the bright-end sampling exceptionally small volumes and behaving in a non-Euclidean manner. An example of incomplete points can be seen at 150\,MHz (see Figure\,\ref{fig:A1}, middle panel), normalized to the Euclidean expectation. In data from the GLEAM survey \citep{Hurley-Walker_2016}, for example, the faintest data points are incomplete and clearly fall below the other data points, and hence flagged as incomplete. The reason for data to be omitted is also highlighted at 610\,MHz, where the brightest three points are shown (see Figure\,\ref{fig:A1}) but again flagged since a turn upwards in the counts at bright fluxes is nonphysical (and likely indicative of a local group). 

In the very local Universe we expect the source counts to follow the Euclidean expectation, i.e., neither any curvature nor the expansion has a noticeable impact over short distances. Hence we also incorporate a prior in our fitting by adding a very bright "anchor-point" representing a Euclidean extrapolation of the brightest data points. In this case we generate an anchor point at $10^{6}$\,mJy and this is shown as the green circle in Figure\,\ref{fig:fig2}. At the best studied frequencies, the anchor point is not really necessary but in some of the modern frequencies where all sky surveys have not been conducted it is necessary to ensure a physical spline fit to the source count data (see for example the 8.4\,GHz where without the anchor point a simple spline fit is going to trend downward rather than stay flat (see middle panel of Figure \ref{fig:A2}). To be consistent, we adopt an anchor point at all frequencies. We also show the frequency-corrected source count from 1.4\,GHz at 22.26\,Jy as confirmation of our adopted anchor point, and to provide further constraint on the known bright-end behaviour of the source count fits.  A fixed value for $\alpha$ of 0.4 was chosen for this purpose.

\subsection{Towards homogeneous errors}
'The data gleaned from the literature often provide errors derived in different ways. Hence some effort is required to homogenise the errors before any weighted fitting of the combined source count data. For a meaningful statistical fit it is also necessary to ensure the errors are realistic. To address this we calculate new errors for all datasets using the combination of a conservative error floor estimate to represent systematic uncertainties, and the Poisson expectation from the source-counts. For each frequency we increase the error-floor until the combined datasets are statistically consistent. This typically resulted in an error floor varying from  3\% to 20\%.  To make the errors fully consistent with 68\% rms, most surveys in Table~\ref{tab:Table_2} needed a 3--6\% in the absolute flux scale, or sometimes 10\%, to make the survey consistent with the other surveys in the same flux range. Some of the 8.4 GHz and 325 MHz surveys needed as much as 20\%. For the 8.4 GHz surveys, this is likely due to the more uncertain absolute flux-scale zero point at the highest interferometer frequencies, but at 325 MHz, it could also be due to instrumental confusion being a more important issue in the older interferometer images. For this reason, we checked in Table~\ref{tab:inputdata} that none of the surveys had their faintest source count data points approach the confusion limit. To calculate this, we added two columns to Table \ref{tab:inputdata}, namely the FWHM of each survey's synthesized beam (in arcsec), and in the last column we list the formal 5$\sigma$ point source detection limit for each survey as well as the resulting confusion source density. The latter was computed as follows: For each survey, we used the upper panel of Figure \ref{fig:fig2} to calculate the \textit{integrated} source density down to the 5$\sigma$ point source detection limits above. Next, we used each survey's synthesized beam FWHM-value --- converted to units of square degrees --- to calculate the number of independent beams available for each detected source down to that surveys' 5$\sigma$ completeness limit. Depending on the actual slope of the source counts, typical levels quoted for the confusion limit are 25--50 independent beams per detected source \citep[see, e.g.,][and references therein]{Kramer_2022}. The inverse number of the confusion source density calculated this way is shown on the top line in the last column of Table~\ref{tab:inputdata}, and the 5$\sigma$ point source detection limit is shown on the second line for each survey. In general, all surveys are well below the confusion limit of 0.02--0.04 detected  
sources per beam. 

An example can be seen at 325 MHz in Figure \ref{fig:A1} where there are small systematic offsets between the data sets in which an error floor of 20\% is needed to bring the contributing datasets into statistical agreement prior to variance weighted spline-fitting. The final values for each error floor for all frequencies are given in Table \ref{tab:Table_2}. The relatively large errors at 5 and 8.4 GHz are due to the very small areas covered by the primary beams of radio interferometers at these frequencies, so that these high frequency surveys likely have a significant Cosmic Variance (CV) component given the few areas covered. The cause of the larger errors at the low frequencies of 325-610 MHz can be one or more of the following, the survey's primary beams here are much larger, so CV is likely much smaller, but the larger synthesized beams may lead to more source confusion, and the lower frequency may include a significant fraction of sources whose spectral shape is not a simple power-law due to, e.g., synchrotron self-absorption in compact radio sources. We leave this topic to future study, and will adopt the errors in Table \ref{tab:Table_2} here as needed for a consistent error-budget, whichever their cause. 

\begin{table}

\caption{Imposed error floors for each frequency used. The error floors were added in quadrature to the existing errors.}
\begin{tabular}{cc}
\cline{1-2}
Frequency (MHz) & Imposed Error Floor (\%)  \\ \hline
150 \fixv            & 6                   \\ 
325  \fixv           & 20                    \\ 
610  \fixv           & 10                  \\ 
1400  \fixv          & 3                   \\ 
3000   \fixv         & 3                    \\ 
5000   \fixv         & 10                   \\ 
8400    \fixv        & 20            \\ \hline
\end{tabular}
\label{tab:Table_2}
\end{table}

\subsection{The Windhorst (2003) compendium}
A summary table for the data compendium originally assembled by \citet{Windhorst_2003} is shown in the Appendix. The data was assembled to provide a summary of the current state of radio source counts at three frequencies as of 2003 \citep[see e.g.][]{Windhorst_1985, Windhorst_1990, Windhorst_1993, Windhorst_2003}, and as an aid to predict the extremely faint source counts in the microjansky and nanojansky regime, which may be observed by the SKA \citet{Hopkins_2000}. Despite its age, this compendium provides the majority of the 5\,GHz and 8.4\,GHz source counts, and still includes the faintest 1.4\,GHz source counts.


\section{The integrated source count measurement and the radio EBL}
The source counts, Euclidean normalised source counts and the contribution to the radio EBL at 3.0\,GHz are displayed in Figure\,2 as upper, middle and lower panels respectively. Focusing on the lower panel which shows the differential contribution to the radio EBL, one sees two peaks at $\sim$$10^{2}$\,mJy and $\sim$$10^{-2}$\,mJy. These represent the contribution to the 3\,GHz EBL dominated by AGN and SFG respectively. Below we describe how we now fit these data via simple spline fitting and derive our integrated EBL measurements for each frequency.

\subsection{Spline fitting the source counts}
For each frequency we fit a smooth spline of order 7 adopting a weighting for each data point inversely proportional to the adopted $1\sigma$ error. Hence points with larger error bars have a smaller weight than those with smaller error bars. For our anchor points we adopt an error of 10\% and we also include a transposed bright source count data point from 1.4\,GHz modified to the target frequency assuming a spectral index of 0.4. Note that the inclusion of an error-floor at each frequency assures that the data is statically consistent. We note that without the adoption of this error floor the spline behaviour becomes erratic as the errors vary significantly between surveys. This suggests that unknown systematics are indeed at work at the level indicated in Table \ref{tab:Table_2}. These are most likely caused by calibration issues between surveys, receivers and teams.

Spline fitting alone serves as a useful tool for integrating the data over the range where data exists, and integrating over this range allows us to obtain a lower limit to the radio EBL. The splines can also be extrapolated and if the source counts have converged, this is reasonable as the flux in the extrapolation is likely to be small (and Monte Carlo simulations will provide robust errors). Hence for all data sets we see that the inclusion of the anchor point and transposed 1.4\,GHz results in bright-end convergence in a manner which is consistent with our physical expectation. In all cases we can also identify a clear dip between the AGN and SFG population. 

\subsection{Determination of the EBL and radio temperature}
\begin{figure}
\vspace*{0.000cm}
\includegraphics[width=\columnwidth]{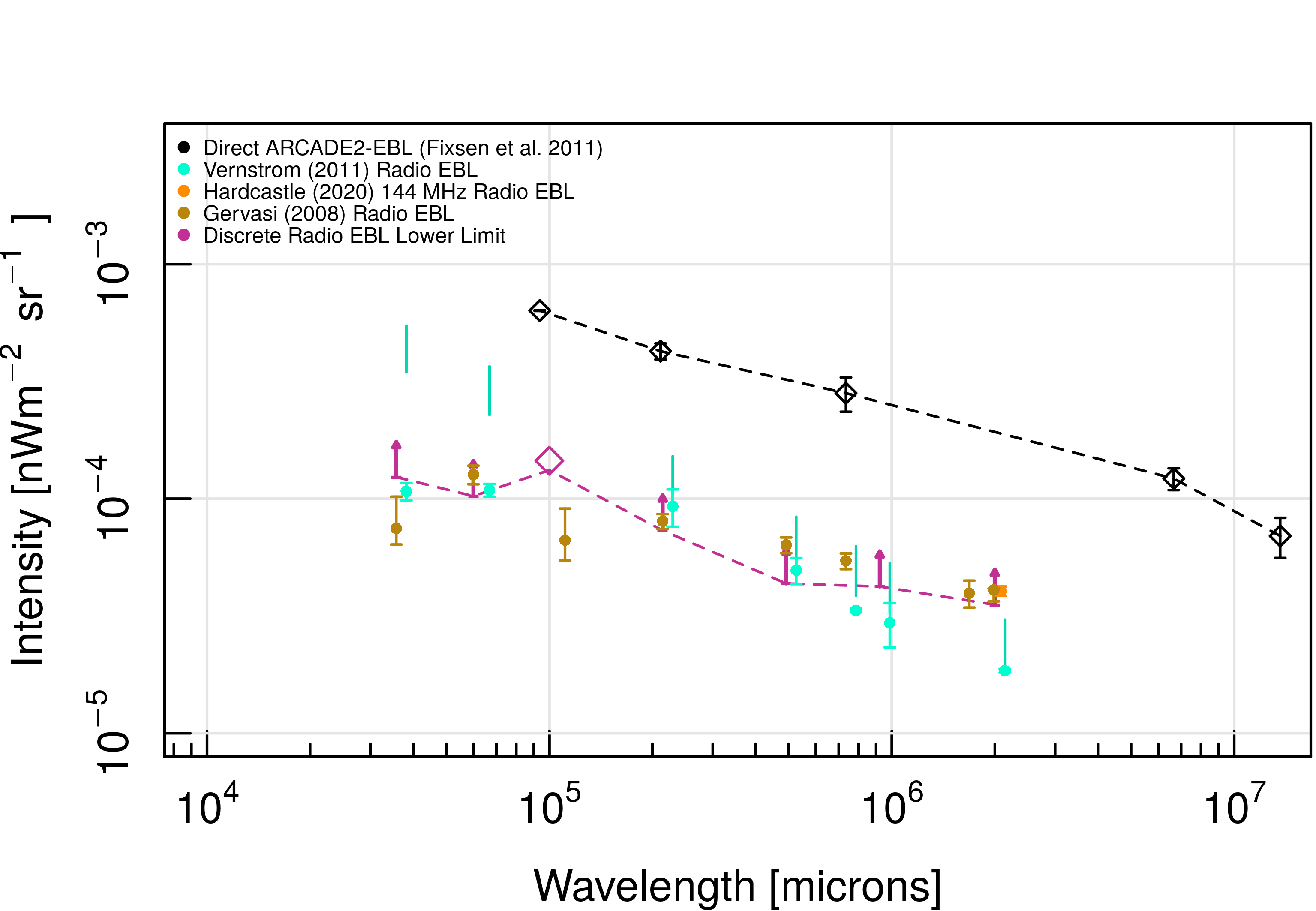}
\caption{A magnified view of the region of interest, highlighting the upper and lower bounds of the radio EBL measurements from ARCADE-2 \citet{Fixsen_2011}. The \citet{Vernstrom_2011} data is shown with the $1-\sigma$ errors as the data points and the range between their minimum/maximum extrapolation as the solid lines. The lower limits are derived from the source counts to the lowest intensity of the surveys, and are shown as the horizontal line at the bottom of the arrow. The convergent measurement, where the integration includes extrapolation of the spline outside the range of the data, is shown as the diamond at 3\,GHz or $10^5$ microns. The error bar for this convergent measurement is too small to meaningfully display, but the $1\sigma$ uncertainty is given in Table \ref{tab:Table_3}}
\label{fig:fig3}
\end{figure}

The total intensity of the radio EBL is determined by integrating the spline fit shown in Figure \ref{fig:fig2} along the x axis. This yields a result in terms of \textit{nW}/m$^{2}$/sr, given by the formula:
\begin{equation}
I=\ln(10) \times \nu \times 10^{-26}\frac{W}{Jy}\times 10^9\frac{nW}{W} \sum (Sp_i(y)+2.0 Sp_i(x)-2.0 \log(10)) {\frac{Jy}{sr}}
\end{equation}
for the spline function fit to the data, labeled Sp. The spline does not have a closed form solution so the integral is shown as a sum, with the necessary unit conversions out front. The results are shown in Figure \ref{fig:fig3}.

The intensity of the EBL given by Equation 2 can be converted to a more conventional brightness temperature via the equation below
\begin{equation}
I=2T \times k_{b} / (\lambda^2 \times \nu)
\label{eqn:eqn2}
\end{equation}

The integral of our limiting fluxes as well as for the extrapolations are shown in Table\,\ref{tab:Table_3}, noting that the extrapolation only converges at 3\,GHz. Errors are determined by taking the 83rd and 17th percentile limits from the series of spline fits for the upper and lower 1$\sigma$ errors shown in the table and figures, respectively.

Figure\,\ref{fig:fig3} shows our measurement of lower limits (maroon diamond or arrows), compared to the previous study of source counts by \citet{Vernstrom_2014} (cyan measurements) and to the direct measurements from ARCADE-2 \citep{Fixsen_2011} (black arrows). In general our lower limits place more stringent constraints on the EBL and help to close the gap somewhat between the previous EBL range and the ARCADE-2 data. However, where we have a clear measurement at 3\,GHz we see that our integrated source counts return an EBL value a factor of $\approx 4\times$ less than the ARCADE-2 measurement. \citet{Seiffert_2011} argue that the discrepancy could be due to any of: underestimated galactic foreground, unaccounted for contribution of background radio sources or some combination of both, though our results show that an undetected population of discrete background sources can not contribute significantly to the discrepancy.

\subsection{Determination of Empirical AGN and SFG Estimates of the CRB}

At this point, and noting that the brighter AGN hump is bounded at all frequencies, we can determine the total AGN-dominated EBL without the need for models as the integral of the spline-fit from $10^5$\,mJy to the crossover point between the AGN and SFG populations. For example, at 3\,GHz the crossover point occurs at just below 1\,mJy, see Figure \ref{fig:fig2}. This crossover point was first identified at a similar flux level at 1.4 GHz by \citet{Windhorst_1985}. We can then define the SFG-dominated EBL contribution as a lower limit by integrating from the crossover point to the faintest data point. The crossover point is determined by finding the flux where the slope between the two peaks changes sign from negative to positive as we move along the spline from bright to faint fluxes. We note that only at 3\,GHz is our estimate of the contribution of the SFG's to the total EBL convergent, due to the noise-analysis data of \citet{Vernstrom_2014}. In all other bands we report lower limits in Table\,\ref{tab:Table_3}. This analysis provides a model independent estimate of the CRB for the two populations, though it is not possible to separate the two populations with source counts alone as both span a wide range in flux. Thus we turn to model source counts in \ref{Section 4} to better understand the CRB.

\begin{table*}
\caption{Intensity of integrated source counts across the data and from and to the estimated midpoint to derive lower limits for the AGN and SFG population EBL.}
\begin{center}
\label{tab:Table_3}
\begin{tabular}{ccccccc} \hline
Frequency & Lower & Integration  & Integration & Integration & Integration  \\ 
& limit & to lower limit & to zero & to Midpoint (AGN) & From Midpoint (SFG) & Estimated Midpoint 
\\\,MHz & $\mu$Jy & nW m$^{-2}$ sr$^{-1}$ & nW m$^{-2}$ sr$^{-1}$ & nW m$^{-2}$ sr$^{-1}$ & Value $\log_{10}($mJy$)$\\ \hline
150& 220      & $3.55^{+0.01}_{-0.01}\times10^{-5}$  &  N/A &  $2.83_{-0.01}^{+0.01} \times 10^{-5}$ & $7.21_{-0.02}^{+0.02} \times 10^{-6}$ & 0.64 \\ 
325& 242     & $4.32^{+0.06}_{-0.05}\times10^{-5}$  & N/A  &  $3.81_{-0.06}^{+0.06} \times 10^{-5}$  & $5.13_{-0.07}^{+0.06} \times 10^{-6}$ & 0.34 \\
610& 56     & $4.42^{+0.06}_{-0.06}\times10^{-5}$  & N/A & $3.21_{-0.05}^{+0.06} \times 10^{-5}$ & $1.21_{-0.01}^{+0.01} \times 10^{-5}$ & 0.28 \\
1400& 12.4     & $7.74^{+0.03}_{-0.03}\times10^{-5}$ & N/A & $5.22_{-0.02}^{+0.02} \times 10^{-5}$ & $2.52_{-0.02}^{+0.02} \times 10^{-5}$ & -0.02 \\ 
3000& 0.05$^\dagger$  & $1.27^{+0.02}_{-0.02}\times10^{-4}$  & $1.31_{-0.04}^{+0.04} \times 10^{-4}$  & $6.74_{-0.02}^{+0.02} \times 10^{-5}$ & $6.29_{-0.41}^{+0.43} \times 10^{-5}$ & -0.11 \\ 
5000 & 18.19      & $1.04^{+0.03}_{-0.03}\times10^{-4}$   & N/A  &   $8.50_{-0.13}^{+0.12} \times 10^{-5}$ & $1.84_{-0.12}^{+0.13} \times 10^{-5}$ & -0.68 \\ 
8400 & 13.4      & $1.23^{+0.04}_{-0.05}\times10^{-4}$   & N/A & $1.03_{-0.05}^{+0.04} \times 10^{-4}$ & $2.54_{-0.16}^{+0.15} \times 10^{-5}$ & -0.74 \\ \hline
\end{tabular}

\end{center}

$^\dagger$ The 3000\,MHz faintest point is from \citet{Vernstrom_2014} $P(D)$ noise analysis. The faintest direct source count measurement at 3000\,MHz is at 3.72\,$\mu$Jy. All other data sets have a faintest direct source count measurement. Column 3 contains the integral from the $10^5\,\mu$Jy point to the faintest data point, and column 5 is the integral of the AGN-dominated sources from the $10^5\,\mu$Jy to the estimated midpoint value. The integration of the AGN-dominated peak of the radio EBL contribution is discussed in Section 4.1.
\end{table*}

\begin{table*}
\renewcommand\thetable{3B}
\caption{The EBL values from Table \ref{tab:Table_3} converted to a brightness temperature in mK, the units used in \citet{Vernstrom_2011}.}
\begin{center}

\begin{tabular}{ccccccc} \hline
 Frequency & Integration  & Integration & Integration & Integration  \\ 
 & to lower limit & to zero & to Midpoint (AGN) & From Midpoint (SFG)  \\\, MHz &  mK & mK & mK & mK \\ \hline
150 & $3.42^{+0.01}_{-0.01}\times10^{4}$  &  N/A &  $2.73^{+0.01}_{-0.01}\times10^{4}$ & $6.95^{+0.02}_{-0.02}\times10^{4}$ \\ 
325 & $4.10^{+0.06}_{-0.05}\times10^{3}$  & N/A  &  $3.61^{+0.06}_{-0.06}\times10^{3}$ & $4.86^{+0.01}_{-0.01}\times10^{3}$  \\
610 & $6.34^{+0.09}_{-0.09}\times10^{2}$  & N/A & $4.60^{+0.09}_{-0.07}\times10^{2}$ & $1.74^{+0.09}_{-0.10}\times10^{2}$  \\
1400 & $91.8^{+0.04}_{-0.04}$ & N/A & $62.0^{+0.24}_{-0.24}$ & $30.0^{+0.24}_{-0.24}$  \\ 
3000 & $15.3^{+0.24}_{-0.24}$  & $15.8^{+0.48}_{-0.48}$  & $8.14^{+0.02}_{-0.02}$ & $7.58^{+0.52}_{-0.49}$  \\ 
5000 & $2.71^{+0.08}_{-0.08}$   & N/A  &   $2.21^{+0.03}_{-0.03}$ & $0.48^{+0.03}_{-0.03}$  \\ 
8400 & $0.675^{+0.02}_{-0.03}$   & N/A & $0.566^{+0.02}_{-0.03}$ & $0.139^{+0.01}_{-0.01}$  \\ \hline
\end{tabular}

\end{center}

\end{table*}

\section{Using SHARK to extract the total EBL}
\label{Section 4}
To obtain a more robust and physically motivated estimate for both the AGN and SFG contributions to the EBL, we need to incorporate a model. In the past this has typically been done by backward modelling local radio luminosity functions for SFGs and adopting some evolution to determine the predicted faint source counts. These are tweaked to match the flux range where data exists and then used to extrapolate to recover the flux beyond the survey limits. Here we adopt a slightly different strategy by using a forward semi-analytic model, built on N-body simulations, where we use the form (shape) of the predicted source counts to help integrate our SFG population.

We elect to use the {\sc Shark} semi-analytic model of galaxy formation \citep{Lagos_2018} which is able to generate self-consistent predictions from Ultraviolet to radio continuum emission due to star formation. {\sc Shark} models a range of baryon physics that are thought to play an important role in the formation and evolution of galaxies, including:  (i) the collapse and merging of dark matter (DM) halos; (ii) the accretion of gas onto halos, which is modulated by the DM accretion rate; (iii) the shock heating and radiative cooling of gas inside DM halos, leading to the formation of galactic discs via conservation of specific angular momentum of the cooling gas; (iii) star formation in galaxy discs; (iv) stellar feedback from the evolving stellar populations; (v) chemical enrichment of stars and gas; (vi) the growth via gas accretion and merging of supermassive black holes; (vii) heating by active galactic nuclei (AGN); (viii) photoionization of the intergalactic medium; (ix) galaxy mergers driven by dynamical friction within common DM halos which can trigger bursts of SF and the formation and/or growth of spheroids; (x) collapse of globally unstable disks that also lead to the bursts of SF and the formation and/or growth of bulges. For this paper, we use the default model presented in \citet{Lagos_2018}.

\citet{Lagos_2019} and \citet{Lagos_2020} presented the FUV-to-FIR multi-wavelength predictions of {\sc Shark}, by combining the model with the generative SED model {\sc ProSpect} \citep{Robotham_2020}, and using a novel method to compute dust attenuation parameters based on the radiative transfer analysis of the EAGLE hydrodynamical simulations of \citet{Trayford_2020}. The model was shown to produce excellent agreement with the observed number counts from the FUV to the FIR \citep{Lagos_2019}, and the redshift distribution of FIR-selected sources \citep{Lagos_2020}. Here, we use the SED model extension of Hansen et al. (in preparation), which models the radio continuum emission of {\sc Shark} galaxies. In particular, we use the predictions based on the implementation of the empirical \citet{Dale_2014} model, which assumes a FIR-radio correlation, and has as free parameters, the ratio of free-free-to-synchrotron emission (set to $0.1$), and the frequency dependence of the free-free and synchrotron SED emission (which are set to be $\nu^{-0.1}$ and $\nu^{-0.8}$, respectively); these are the values used by \citet{Dale_2014}. To compare with our observed radio wavelength number counts, we use the lightcone introduced in \citet{Lagos_2019}, Section 5, which has an area of $107$~deg$^2$, and includes all galaxies with a dummy magnitude, computed assuming a stellar mass-to-light ratio of $1$, $<32$ and at $0\le z\le 8$. The method used to construct lightcones is described in \citet{Chauhan_2019}. In this version of the SHARK SEDs, we do not include AGN radio emission, but see \citet{Amarantidis_2019} for a comparison of the SHARK 1.4\,GHz luminosity function of AGN with observations over a wide redshift range.
The use of the SHARK model lightcones and their derived source counts for the SFG population only at all seven frequencies allows us to obtain a model radio EBL contribution. However, while the models come reasonably close they do not precisely fit the observed source counts with some small amplitude offsets. In order to better match the data and models, we derive a scaling factor which simply shifts the entire model source count fit up or down by a constant factor to fit to the available data. This scaling factor is obtained from a chi-squared minimization between the relevant data points that sample the SFG peak and the model. The fit is weighted by the inverse error of each data point, i.e., we minimise $\sum_i \frac{(D_i-M_i)^2}{\sigma_i}$ where $D$ represents the data point, $M$ the model and $1\sigma$ the error on the data-point. Here we need to be careful to only include data points dominated by star-formation and not contaminated by significant AGN flux. Hence we only use data points with fluxes below where the the minimum between the two peaks occurs and down to the flux limit. The scaling we derive to the SHARK number counts is small, only reaching $13\%$ in either direction. SHARK is prone to cosmic variance effects and we sample a simulated volume of 310 $Mpc^3$. The source fluxes are not prone to random error, but have systematic errors associated with the choice of model input parameters. Thus, the correction to match observed data can be done in either direction. Table~\ref{tab:Table_4} shows the upper and low flux limits used to determine the scaling factors in each frequency. We derive errors for this process by letting the data points used in this minimization vary within their allowed errors which returns a different scale factor each time from which we pick a best fit, upper, and lower limit from the median, $83^{rd}$, and $17^{th}$ percentile of the resulting distribution. 

\begin{table*}

\caption{Derived factors by which the SHARK model EBL curves were scaled to match the data. The best scale factor was determined to be the median of the distribution of scale factor values, while the upper and lower limits are the 83rd and 17th percentiles, respectively.}
\begin{tabular}{ccccc}
\cline{1-5}
Frequency (MHz) & Flux range  & SHARK Model  & Scale Factor  & Scale Factor \\ (MHz) & ($\log_{10}($mJy$)$) & Scaling Factor & Upper Limit & Lower Limit  \\ \hline
150 \fixv            & $<0.4$              & 0.906         & 0.912         & 0.900                    \\ 
325  \fixv           & $<0.4$               & 0.876          & 0.984        & 0.857                    \\ 
610  \fixv           & $<0.1$              & 0.943         & 0.951         & 0.934                   \\ 
1400  \fixv          & $<-1$              & 0.998         & 1.010         & 0.985                  \\ 
3000   \fixv         & $<-1$              & 0.974         & 0.987         & 0.960                    \\ 
5000   \fixv         & $<-1$               & 1.034          & 1.087        & 0.983                    \\ 
8400    \fixv        & $<-1$            & 1.136          & 1.208         & 1.044                    \\ \hline
\end{tabular}
\label{tab:Table_4}
\end{table*}

\subsection{Sub-dividing the EBL into the AGN and SFG contributions}
\begin{figure*}
\includegraphics[width=\textwidth]{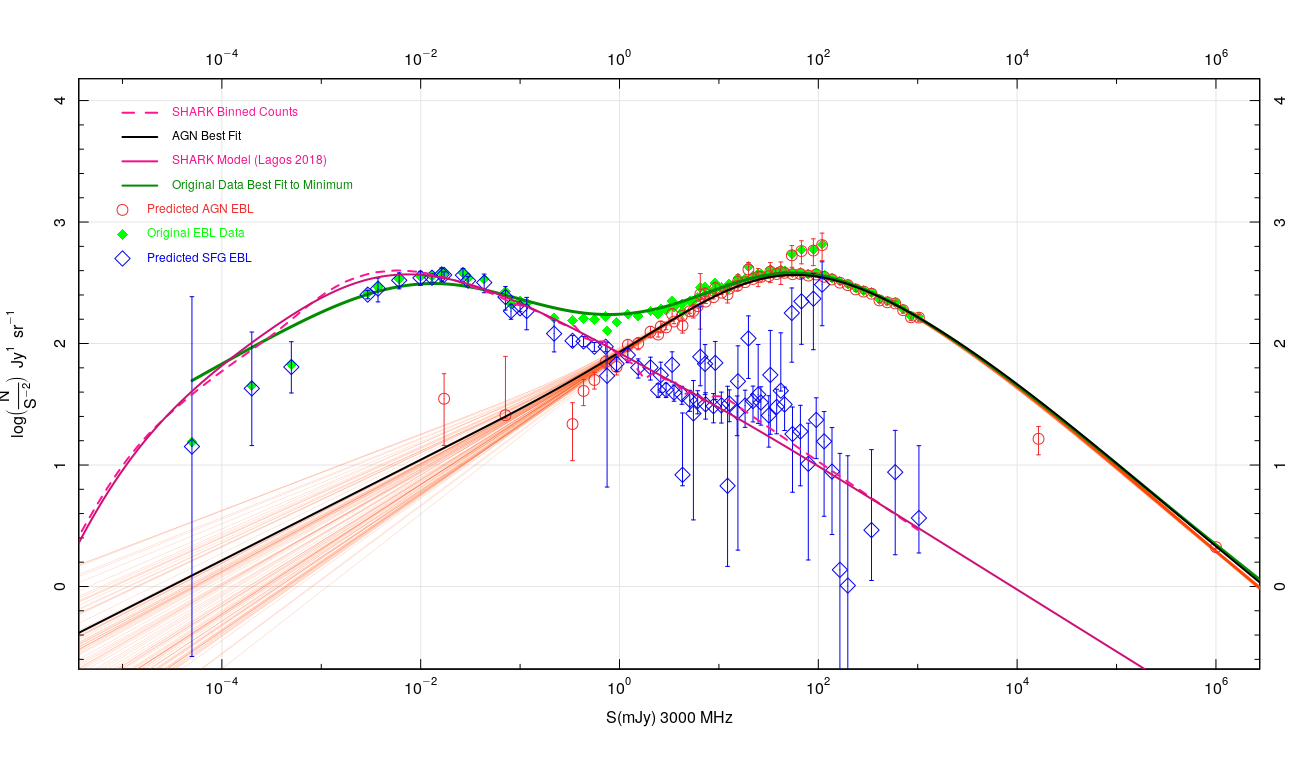}
\caption{A breakdown of the 3\,GHz source counts into its respective populations. The 1.4\,GHz anchor point and Euclidean extension are shown and used in the fit for the predicted AGN EBL, and the SHARK model is given a constraining point shown at $10^3$\,mJy to impose Euclidean behavior. 
\label{fig:fig4}}
\end{figure*}

Using the data normalised SHARK model predictions, we are now able to subtract the SFG population only from the original source count data.  In doing so we obtain a representation of the AGN source counts only and hence can derive the separable contributions of the SFG and AGN to the total EBL. Note that although a model has been used, the AGN source counts are minimally dependent on the SFG model because the point which overlaps with the AGN peak is generally Euclidean and hence predictable in behaviour. Figure\,\ref{fig:fig4} illustrates the process at 3\,GHz. This shows the original combined source count data as green diamonds and the normalised SHARK prediction for the SFG as the magenta line. In order to generate a model value at the location of every data point we fit the model with a spline and then subtract this value from the counts. What remains is the data associated with AGN (red data points), which we can now subtract from the original source counts to get the data for the SFG only (blue data points). As expected, the right peak of the radio EBL seen in Figure \ref{fig:fig4} is almost entirely dominated by the AGN population, and the left peak is dominated by the much fainter SFG population. It is also reassuring to see that the blue points scatter around the SHARK model even over the flux ranges not used in the deriving the scaling factors.

We obtain error measurements by 1000 samplings of the scale factors assuming a Poisson distribution as represented by the values shown in Table \ref{tab:Table_4}, and also allowing all data points to vary independently according to their assigned errors. Hence for each iteration we obtain a randomly sampled measurement of the SFG contribution from the renormalised SHARK model counts alone, and the AGN contribution from the original data with the SFG prediction removed. Using 1000 different spline models each with a different scale factor and data adjusted according to the errors by a normal distribution, we obtain a set of 1000 EBL measurements for both the AGN and SFG populations, and their summed total as well. While MCMC fitting works well for the $\alpha$ vs flux relationship, the complexity of the data and the lack of an analytical solution means that MCMC analysis is not viable for fitting the EBL distribution shown in Figure \ref{fig:fig4}.

\begin{figure*}
\includegraphics[width=\textwidth]{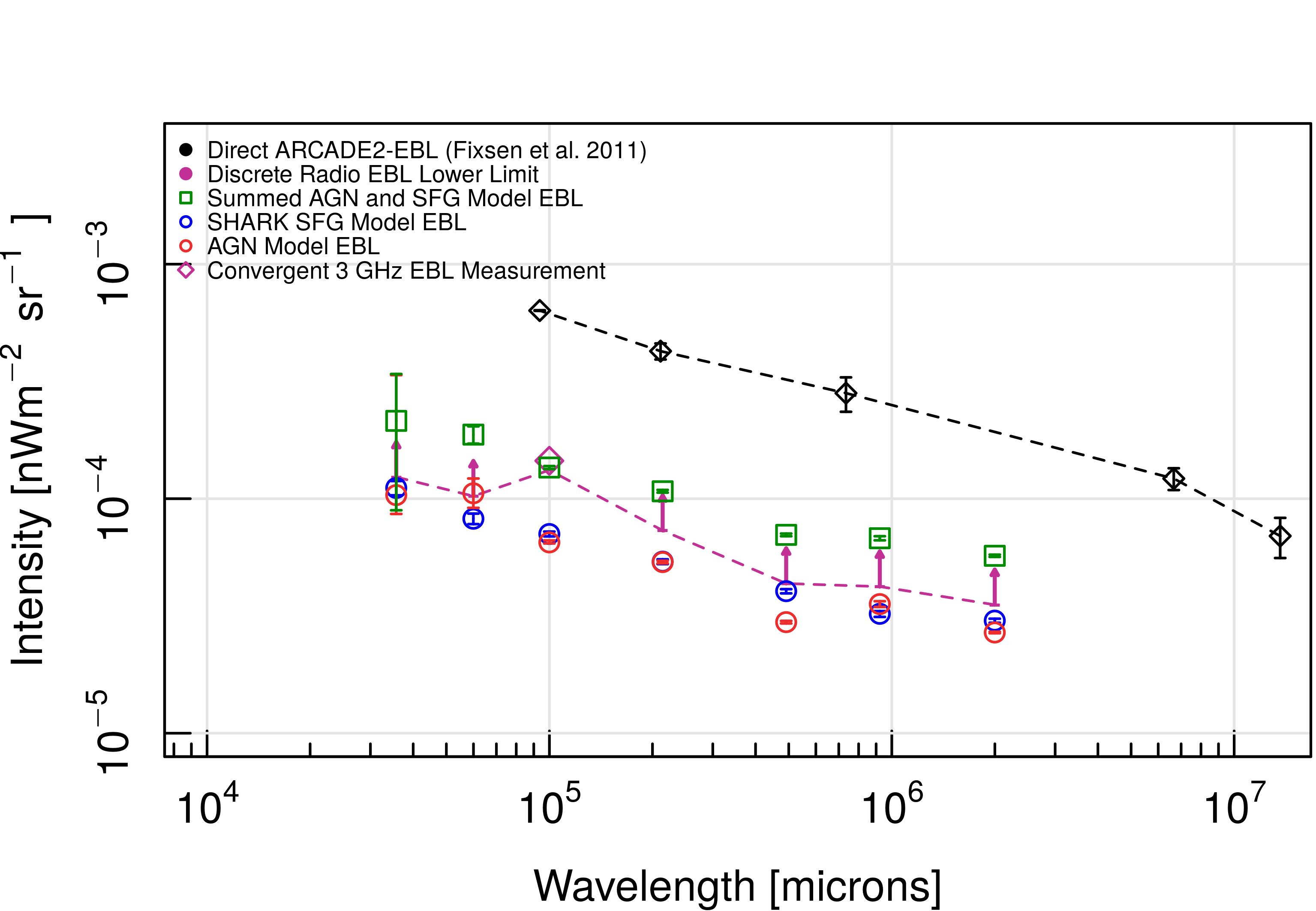}
\caption{ The model-derived estimates of the source counts shown alongside the lower limit estimates and convergent measurement at 3\,GHz. The EBL is broken down into its two dominant components and the sum of the two model-derived components. A dashed line joining the original lower limits is displayed, and indicates that in all frequencies the lower limit measurements still contain contributions from both populations, and that at 3\,GHz, their sum is consistent with the only convergent measurement.
\label{fig:fig5}}
\end{figure*}

In all 3 cases, we use the same method of taking the median, $83^{rd}$, and $17^{th}$ percentiles to obtain a measurement, upper, and lower limit, which are displayed in Figure \ref{fig:fig5} with their associated errors (blue points represent the SFG, red points the AGN and green squares the total EBL). Also shown are the ARCADE-2 direct measurements of the total radio EBL (black line and arrows) and our previous total lower limits from spline fitting only (magenta line arrows and diamonds). In these figures the previous data from \citet{Vernstrom_2011} shown on Figure\,\ref{fig:fig3} is left off for clarity. Figure\,\ref{fig:fig5} allows us to compare our model dependent extrapolation of the total discrete radio EBL at all frequencies, and compare to our lower limits and convergent measurement at 3\,GHz. At 3\,GHz, the sum of the two populations agrees extremely well with the empirical measurement to within $\approx7\%$. This is consistent with our derived error from our spline fit of $\pm 9$ \%. Given the different parameters required to generate the SHARK model and large errors on the convergent source counts at this frequency, the consistency of this adds credence to our methodology that separates the discrete EBL into its two primary components: SFG and AGN. The radio EBL contribution for each population, errors and their combination is shown in Table \,\ref{tab:Table_5}.

\begin{table*}
\caption{Summary of integrated radio source count data at different frequencies. Values here are quoted with upper and lower error bars, and the original lower limits are re-stated for comparison to the model-derived values. All EBL values quoted are given in units of nW m$^{-2}$ sr$^{-1}$.}
\begin{tabular}{cccccc}
\hline
Frequency (MHz) & SHARK Model EBL & AGN-Fit EBL & Total Sum  & Original Lower Limit \\ \hline
150 \fixv          & $3.02^{+0.06}_{-0.06}\times10^{-5}$               & $2.68^{+0.02}_{-0.03}\times10^{-5}$           & $5.71^{+0.06}_{-0.06}\times10^{-5}$          & $3.53^{+0.03}_{-0.03}\times10^{-5}$    \\ 
325 \fixv            & $3.23^{+0.09}_{-0.09}\times10^{-5}$               & $3.55^{+0.10}_{-0.10}\times10^{-5}$           & $6.74^{+0.13}_{-0.14}\times10^{-5}$         & $4.23^{+0.24}_{-0.21}\times10^{-5}$            \\ 
610 \fixv            & $4.03^{+0.09}_{-0.08}\times10^{-5}$               & $2.98^{+0.04}_{-0.04}\times10^{-5}$           & $7.01^{+0.09}_{-0.08}\times10^{-5}$         & $4.34^{+0.04}_{-0.04}\times10^{-5}$     \\ 
1400 \fixv            & $5.39^{+0.13}_{-0.12}\times10^{-5}$               & $5.37^{+0.07}_{-0.07}\times10^{-5}$           & $1.08^{+0.01}_{-0.01}\times10^{-4}$         & $7.3^{+0.06}_{-0.07}\times10^{-5}$     \\ 
3000 \fixv            & $7.07^{+0.17}_{-0.16}\times10^{-5}$               & $6.51^{+0.13}_{-0.67}\times10^{-5}$           & $1.36^{+0.02}_{-0.02}\times10^{-4}$         & $1.32^{+0.05}_{-0.04}\times10^{-4}$     \\ 
5000    \fixv        & $8.21^{+0.42}_{-0.42}\times10^{-5}$               & $1.05^{+0.16}_{-0.14}\times10^{-4}$           & $1.86^{+0.21}_{-0.15}\times10^{-4}$         & $1.03^{+0.04}_{-0.04}\times10^{-4}$     \\ 
8400 \fixv            & $1.11^{+0.08}_{-0.08}\times10^{-4}$               & $1.03^{+2.33}_{-0.175}\times10^{-4}$           & $2.19^{+2.29}_{-0.244}\times10^{-4}$         & $1.23^{+0.989}_{-0.107}\times10^{-4}$     \\ \hline
\end{tabular}
\label{tab:Table_5}
\end{table*}

\begin{table*}
\renewcommand\thetable{6B}
\caption{The EBL values from Table \ref{tab:Table_5} converted to a brightness temperature in mK, the units used in \citet{Vernstrom_2011}.}
\begin{center}

\begin{tabular}{cccccc}
\hline
Frequency (MHz) & SHARK Model EBL & AGN-Fit EBL & Total Sum  & Original Lower Limit \\ \hline
150 & $2.91^{+0.01}_{-0.01}\times10^{4}$  &  $2.58^{+0.02}_{-0.03}\times10^{4}$ &  $5.51^{+0.06}_{-0.06}\times10^{4}$ & $3.42^{+0.01}_{-0.01}\times10^{4}$ \\ 
325 & $3.06^{+0.09}_{-0.09}\times10^{3}$  & $3.36^{+0.09}_{-0.09}\times10^{3}$   &  $6.39^{+0.12}_{-0.13}\times10^{3}$ & $4.10^{+0.06}_{-0.05}\times10^{3}$  \\
610 & $5.78^{+0.13}_{-0.11}\times10^{2}$  & $4.27^{+0.06}_{-0.06}\times10^{2}$ & $1.01^{+0.01}_{-0.01}\times10^{3}$ & $6.34^{+0.09}_{-0.09}\times10^{2}$  \\
1400 & $63.9^{+1.54}_{-1.42}$ & $63.7^{+0.83}_{-0.83}$ & $1.28^{+0.01}_{-0.01}\times10^{2}$ & $91.8^{+0.04}_{-0.04}$  \\ 
3000 & $8.52^{+0.20}_{-0.19}$  & $7.85^{+0.16}_{-0.81}$  & $16.4^{+0.24}_{-0.24}$ & $15.3^{+0.24}_{-0.24}$ \\ 
5000 & $2.14^{+0.11}_{-0.11}$   & $2.73^{+0.42}_{-0.36}$  &   $4.84^{+0.11}_{-0.11}$ & $2.71^{+0.08}_{-0.08}$  \\ 
8400 & $0.610^{+0.04}_{-0.04}$   & $0.566^{+1.28}_{-0.10}$ & $1.20^{+1.26}_{-0.13}$ & $0.675^{+0.02}_{-0.03}$ \\ \hline
\end{tabular}

\end{center}

\end{table*}

\subsection{Comparison to previous source count based EBL measurements}
We compare our results to previous discrete radio EBL measurements such as \citet{Vernstrom_2011}, \citet{Gervasi_2008}, and 
\citet{Hardcastle_2021}. Our measurements and lower limits lie within the associated errors of \citet{Vernstrom_2011}, with the exception of 150\,MHz. The availability of data from the decade since their work has allowed us to move the lower limits of the radio EBL upwards significantly at all frequencies other than 1.4\,GHz
due to the increased availability of deep survey data at these less studied frequencies. For example, previous surveys at 150\,MHz had only achieved a depth of $\approx 106$\,mJy, while data published in this decade now reaches a depth of 220\,$\mu$Jy. A different behavior is seen at 325\,MHz where the newer data sets achieve roughly the same depth but are systematically above the data used in \citet{Vernstrom_2011}. From the results obtained in Section 5.1 and shown in Figure \ref{fig:fig5}, we expect the radio EBL to gradually increase with frequency. Where this is not shown by the data is likely due to the data not reaching faint enough fluxes to become convergent or have a well-defined lower limit, and future surveys reaching fainter fluxes at all frequencies with better statistics should address this inconsistency. At 150 MHz synchrotron self-absorption can become important in shaping the counts of compact sources with a high free electron density. See \citet{Mancuso_2017} for a detailed discussion on how synchrotron self-absorption affects the number counts at 150 MHz.

\subsection{Comparison to direct EBL measurements}
At 3\,GHz, we still do not measure enough flux with convergent source counts to achieve consistency with the ARCADE-2 measurements \citep{Fixsen_2011}. The same is true for our lower limits and model-derived measurements, which also lie below the ARCADE-2 measurement. At 3\,GHz where a convergent measurement is possible, the corresponding ARCADE-2 measurement lies a factor of $\approx 4$ above the discrete measurement obtained from source counts. 

With convergent models such as ones presented here and with ever deeper counts the possibility that the whole of the ARCADE-2 temperature could come from galaxies alone is increasingly unlikely. There are still models for counts below $1\, \mu$Jy, such as those presented in \citet{Hopkins_2000}, \citet{Condon_2012, Vernstrom_2014} that would be enough to account for the difference. However, there would be constraints on the clustering of such sources from radio experiments measuring the angular power spectrum \citep[e.g][]{Holder_2014, Offringa_2022}, which imply either faint sources that are very clustered or sources of high redshift.

Many other possibilities have been looked at for what could make up the difference between the source counts and ARCADE-2 \citep[see][ for an overview]{Singal_2018}. One possibility is diffuse radio emission associated with large-scale structure, or the clusters and the cosmic web. From statistical studies it appears that this sort of diffuse low-level emission may account for a few mK at $1.4\,$GHz \citep{Vernstrom_2014_B,Vernstrom_2017,Vernstrom_2021}. \citet{Krause_2021} examined the local bubble as a possible contributor to the radio background but found at most this would be at the percent level. There has also been a good deal of work looking at the possibility of synchrotron emission from dark matter decay and or annihilation \citep[e.g.][]{Fornengo_2011,Fornengo_2014}, as well as other more exotic explanations \citep[such as dense nuggets of quarks][]{Lawson_2013}. There is also the possibility of a Galactic origin. Certain models that include a spherical halo component for the Galaxy claim to account for any excess \citep{Subrahmanyan_2013}, though there are possible problems with such models \citep[see e.g.][]{Singal_2015}.  

It seems unlikely for the discrepancy to be caused by just one source. At this point it seems more likely that many factors may contribute such as faint clustered point sources, low-level emission from the cosmic web, the local bubble, possibly dark matter annihilation or fast radio bursts, and possible reasons to decrease the ARCADE-2 estimate like revised Galactic modelling. Some of these contributors are easier to investigate than others, but new estimates for many will be coming in the next few years. Radio surveys at a variety of frequencies from the likes of MeerKAT, LOFAR, ASKAP and others will continue to push the source counts and also investigate the clustering properties of faint sources. With these surveys as well better estimates of the contributions from diffuse emission (aided by stacking and other statistical estimates), and continuing to put new constraints on the dark matter models \citep[e.g.][]{Regis_2021}. New deep polarisation surveys like POSSUM will provide many new probes of the Galaxy's magnetic field, particularly in the halo region. However, one important future measurement would be a new measure of the radio sky from experiments with absolute zero-level calibration. 

\section{Predicting the SFG radio EBL from the cosmic star-formation history}
In this section we attempt to predict the radio continuum flux from the SFG population. We start with the Cosmic Star-Formation History (e.g., \citealt{Madau_2014}; \citealt{Driver_2018}) and what we know about the radio-star-formation correlation (\citealt{van_der_Kruit_1971}; \citealt{Condon_1992}; \citealt{Davies_2017}; \citealt{Delvecchio_2021};  \citealt{Seymour_2009}). Star-formation results in strong UV fluxes that ionise the surrounding gas. As massive (M$>8$M$_{\odot}$) stars reach their endpoints they generate significant quantities of charged particles through supernova shocks which interact in the Galactic magnetic field. In addition there is a free-free component that arises from the interaction of less relativistic particles with the Inter Stellar Medium (ISM). Hence there is a direct connection between the supernova rate and radio flux. This leads to a well known correlation between star-formation as evidenced by UV fluxes and optical emission lines, far-IR radiation from heated dust, and radio emission from synchotron radiation around massive stars and SN plus general free-free emission from the interstellar medium (ISM). Critical physical parameters are the temperature of the ISM and the electron density. 

These optical-FIR-radio correlation arising from star-formation has been known for some time and over the past decade enormous amounts of work have been invested in constructing the cosmic star-formation history (CSFH) from the era of recombination to the present day. Here we adopt the EBL model described in \citep{Koushan_2021} in which the spectral energy distribution code {\sc ProSpect} (\citealt{Robotham_2020}), uses a CSFH to provide a prediction of the UV to radio EBL. The detailed modelling of the radio continuum aspect follows closely that described in \citep{Marvil_2015} and is essentially built in to the {\sc ProSpect} code (see Thorne et al, submitted).

\begin{figure*}
\includegraphics[width=\textwidth]{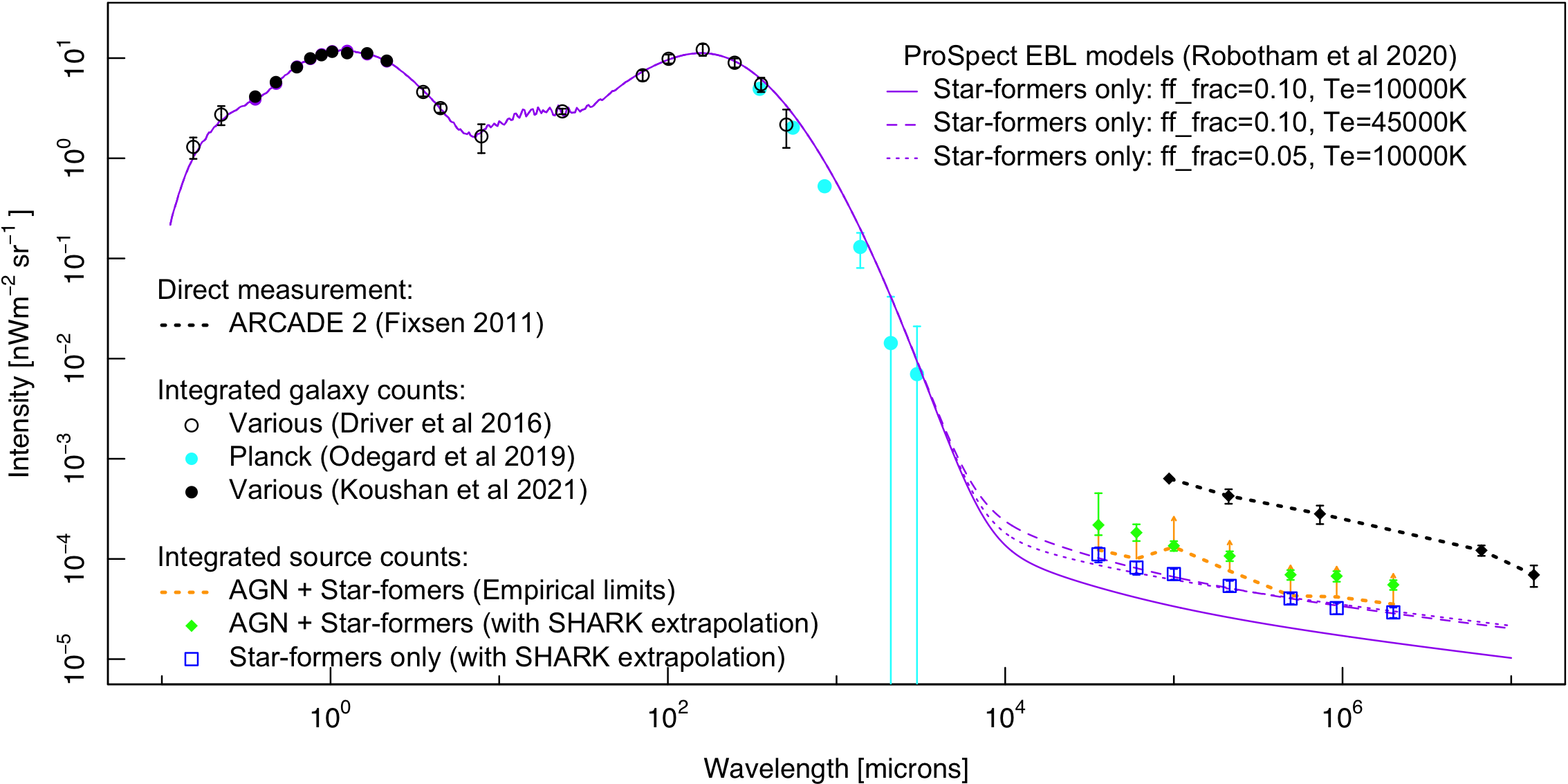}
\caption{The EBL from optical to radio wavelengths showing our new radio EBL empirical lower limits (gold arrows), our extrapolated measurements (green), and the component arising from SFG only (blue squares). Also show is the EBL model predicted using the {\sc ProSpect} code, where the radio emission of the SFG population is predicted entirely from the Cosmic Star-Formation History \citep{Madau_2014}. The initial (default) {\sc ProSpect} model (purple solid line; see \citealt{Robotham_2020}) under-predicts the SFG emission at radio wavelengths. To rectify this one can either increase the ISM temperature (from $10^5$K to $4.5\times10^5$K; dashed purple line), or by changing the balance between the synchrotron and free-free emission (dotted purple line). 
\label{fig:fig6}}
\end{figure*}

Figure\,\ref{fig:fig6} shows the recent compendium of EBL data from a range of optical to infrared surveys compiled over recent years and described in \citet{Driver_2016} and \citet{Koushan_2021}. These compendiums are based on galaxy number-counts (analogous to radio source counts) where in most cases the contribution to the EBL is clearly bounded and the measurements and errors known to an accuracy of 10-20\%. Also shown on Figure\,\ref{fig:fig6} (as cyan data points) are the Planck data from \citep{Odegard_2019} covering the mm regime. Our new data is added at the radio wavelengths as blue open squares showing the measured EBL contribution from the SFG. For completeness we also show the lower limits and measurements of the {\it combined} SF and AGN EBL. The ARCADE2 measurements are also shown.

The model curve (solid purple line) shows our prediction for the \citet{Madau_2014} compendium of the cosmic star-formation history using the default options for {\sc ProSpect}. The model fits the UV to far-IR data extremely well, as noted in \citet{Koushan_2021}, and very slightly over-predicts the \citet{Odegard_2019} data, it also under-predicts the SFG radio EBL quite significantly.

The {\sc ProSpect} model uses the \citet{Marvil_2015} templates for dust and radio emission  which ultimately originate back to the model laid down by \citet{Condon_1992}. Within the {\sc ProSpect} code we have a number of options. The two most relevant are to increase the temperature of the ionised gas, $T_e$, or to modify the fraction of emission which comes from free-free emission rather than synchrotron, {\sc ff\_frac}. In the default model these are set at $T_e=10,000$K (the value proposed by \citet{Condon_1992} based on constraints from the Milky Way. A much better fit can be found by either increasing $T_e$ to $4.5 \times 10^4$K (dashed purple line), or decreasing {\sc ff\_frac} to 0.05 (dotted purple line). We note that \citet{Condon_1990} advocates for the {\sc ff\_frac} to remain in the bounds of $0.05-0.2$.  Both modifications to the {\sc ProSpect}-model provide satisfactory fits to the data and although the exact predictions are slightly different the errors at this stage are too large to distinguish between them. Hence either modification, or some combination thereof, can readily match the data.

 In coming years significant mprovement in the measurement and modelling of the EBL are to be expected as deeper source counts will become available with the advent of the Square Kilometer Array as well as improved understanding of total spectral energy distribution as we assemble large samples of UV-radio SEDs for the normal galaxy population. In due course we might expect the errors in the radio-EBL to become reduced to a few percent at which point it should become possible to extract meaningful constraints on the model parameters.

 Finally, it is worth noting that significant gap between the ARCADE2 data (black dots) and our results on Fig.~\ref{fig:fig6}. While an upward adjustment of the model is possible by increasing the ISM Temperature or free-free fraction further the parameters would need to move out of the recommended bounds to very extreme and physically unlikely values. Adding an additional as yet unknown hidden populations also becomes problematic as such a population could not be regular SFG without very quickly violating the constraints at optical/infrared wavelengths.

\begin{figure*}
\vspace*{0.000cm}
\includegraphics[width=\textwidth]{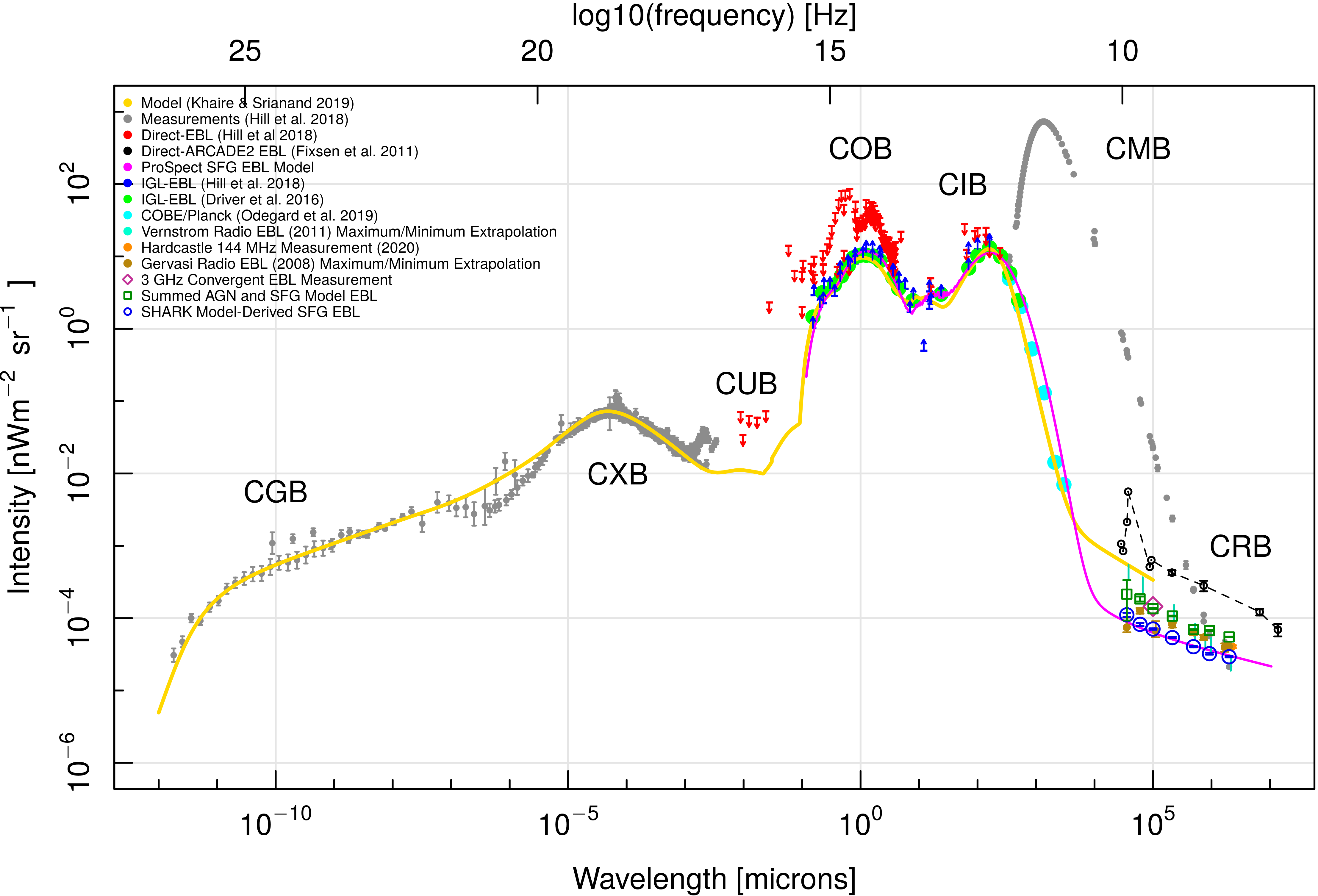}
\caption{The complete EBL over the entire measured EM range, including our discrete radio source count measurements along with those from the literature (as indicated in the Legend). Also shown is the CMB contribution which dominates at most radio wavelengths and the constraints on the total non-CMB radio EBL from ARCADE2 (black dotted line and open circles). Our model-derived (extrapolated) data points in the radio region are shown as green squares (AGN and SFG), and blue circles (SFG). The solid purple line is the best fitting (dotted) SFG only prediction from Figure \ref{fig:fig6}.}
\label{fig:figEBL}
\end{figure*}

\section{Summary} 
In this paper, we measure the discrete radio EBL across 7 frequencies using source count data across nearly 50 years of study and 8 orders of magnitude in flux data. We assembled our data into a massive compendium which we use to establish the counts at each frequency using the spectral index versus flux relationship detailed in \citep[][and references therein]{Windhorst_2003}. We use the data to identify problematic data sets which sit apart from the rest, and to identify individual data points which display incompleteness or non-Euclidean behavior, and establish an anchor point to bring the radio EBL to convergence at the bright end in excess of $\approx 10^4 mJy$. Finally, in preparing the data, we add a conservative error floor estimate for each frequency to address systematic errors in the absolute flux density scale seen between different surveys. We integrate under a series of spline fits to establish lower limits and a rough estimate for the AGN contribution by integrating to the crossover point seen in the radio EBL distribution between the AGN and star-forming galaxy dominated populations. At 3\,GHz, data reaches faint enough fluxes to be convergent at both ends, and we can establish a model-independent convergent measurement of the radio EBL and a separation of its components. To validate our estimates and allow for a convergent measurement at other frequencies, we introduce the SHARK model source counts consisting of only star-forming galaxies, which we use to extrapolate the radio EBL at the faint end of all frequencies and robustly break it down into its respective components dominated by AGN and star-forming galaxies. We compare the model-derived measurements to the data-derived estimates at all frequencies, with special attention at 3\,GHz, where because the data is convergent, we can compare the model source counts to the data, and find a $\approx7\%$ difference between the two. Finally, we establish our model-derived measurements and empirical measurement at 3\,GHz in context with previous studies of the discrete radio EBL such as \citet{Vernstrom_2011} and the EBL as a whole in Figure \ref{fig:figEBL}. We still do not find enough flux in faint resolved sources at 3\,GHz to account for the total flux received by the direct measurement experiment of ARCADE-2 \citep{Fixsen_2011} and do not have a good explanation as to why this is the case.

\section*{Acknowledgement}

We would like to thank the ARC Future Fellowship (FT200100375) on behalf of the University of Western Australia (UWA), and the NASA JWST Interdisciplinary Scientist grants NAG5-12460, NNX14AN10G and 80NSSC18K0200 from GSFC on behalf of Rogier Windhorst and Arizona State University (ASU), both of which provided funding for student
work on this project.

Author Claudia Lagos has received funding from the ARC Centre of 
Excellence for All Sky Astrophysics in 3 Dimensions (ASTRO 3D), through project number CE170100013.

We also are grateful to Yjan Gordon from the University of Wisconsin, Madison for providing the tabulated early VLA Sky Survey source count data shown in \citep{Gordon_2021}. This work was supported by resources provided by The Pawsey Supercomputing Centre with funding from the
Australian Government and the Government of Western Australia.

\section*{Data Availability}
All data has been compiled from the literature as indicated in the Table \ref{tab:inputdata} and Table \ref{tab:Table_5}, and is also made available through a machine readable table available from the MNRAS site.

\bibliographystyle{mnras}
\bibliography{scott.bib}

\section{Appendix}
\begin{table*}     
\renewcommand\thetable{A.1} 
\caption{Summary and breakdown for the \citep{Windhorst_2003} collection. Coordinates given for papers before 2000 are as measured in J1950 and due to the age of some of the work. Missing entries are present for PhD theses which could not be tracked down.}
\begin{tabular}{p{1.0cm}p{3.0cm}p{1.5cm}p{2.0cm}p{3.0cm}p{1.5cm}p{1.5cm}p{2.0cm}} \\ \hline
Frequency  & Reference & Instrument & Field/Survey  & Field Location & Area & Beam FWHM & Confusion Source Density  \\ 
(MHz) & & & Name & & (Ster) & (Arcsec)  & Faintest Source Bin (mJy) \\ \hline
        1400 & \citet{Fomalont_1974} & NRAO 300-ft Telescope & BDFL & Large-Area Survey & 10.22 & 589.31 & $1.01\times 10^{-3}$ \newline 7933 \newline\\ 
        1400 & \citet{Fomalont_1974} & NRAO 300-ft Telescope & BDFL & Large-Area Survey & 4.3 & 589.31 & $9.29\times 10^{-3}$ \newline 2262 \newline\\
        5000 & \citet{Davis_1971} & NRAO 300-ft Telescope & N/A & N/A & $5.79\times 10^{-2}$ & 165.12 & $5.41\times 10^{-4}$ \newline 625 \newline\\ 
        1415 & \citet{Kraus_1972} & Ohio RT & Ohio Sky Survey & Large-Area Survey & 6 & 1006 & $8.10\times 10^{-3}$ \newline 279 \newline\\
        1400 & \citet{Maslowski_1973} & NRAO 300-ft Telescope & Green-Bank Survey &  $\alpha = 11^h 50^m 00^s,\delta = +48^{\circ} 51^{'} 00^{''}$ & 0.159 & 589.31 & $8.69\times 10^{-3}$ \newline 198 \newline\\ 
        1400 & \citet{Machalski_1978} & NRAO 300-ft Telescope & Green-Bank Survey 2 &  $\alpha = 12^h 03^m 00^s,\delta = +36^{\circ} 0^{'} 00^{''}$ & 0.283 & 589.31 & $2.12\times 10^{-2}$ \newline 93.7 \newline\\ 
        1407 & \citet{Pearson_1978} & Cambridge 1-Mile Telescope & Unnamed & 2 Centers & $2.39\times 10^{-4}$ & 23 & $4.92\times 10^{-4}$ \newline  3.53 \newline\\ 
        1415 & \citet{Oosterbaan_1978} & WSRT & Westerbork Background Survey & N/A & $2.57\times 10^{-2}$ & 23 & $3.36\times 10^{-4}$ \newline  6.13 \newline\\ 
        1415 & \citet{Katgert_1974} & WSRT & Unnamed &  ~$\alpha = 01^h 03^m 00^s,\delta = +29^{\circ} 0^{'} 00^{''}$ & $7.62\times 10^{-3}$ & 22 & $2.00\times 10^{-4}$ \newline 11 \newline\\
        1415 & \citet{Le_Poole_1978} & WSRT & Unnamed & N/A & $3.05\times 10^{-4}$ & 23 & $3.98\times 10^{-4}$ \newline 2.13 \newline\\ 
        1412 & \citet{Katgert-Merkelijn_1985} & WSRT & Unnamed & 4 Pointings & $2.59\times 10^{-3}$ & 21 & $6.58\times 10^{-4}$ \newline 1.85 \newline\\ 
        1412 & \citet{Windhorst_1984} & WSRT & Several Fields & Multiple Centers & $1.68\times 10^{-3}$ & 19.79 & $1.78\times 10^{-3}$ \newline 1.31 \newline\\ 
        1412 & \citet{Oort_1985} & WSRT & Lynx 2 &  $\alpha = 08^h 41^m 46^s,\delta = +44^{\circ} 46^{'} 50^{''}$ & $2.60\times 10^{-4}$ & 12.2 & $1.06\times 10^{-4}$ \newline 0.446 \newline\\ 
        1462 & \citet{Windhorst_1985} & VLA & Lynx 2 &  $\alpha = 08^h 41^m 46^s,\delta = +44^{\circ} 46^{'} 50^{''}$ & $2.60\times 10^{-4}$ & 14.33 & $1.78\times 10^{-3}$ \newline 0.294 \newline\\ 
        1412 & \citet{Oort_1987} & WSRT & Lynx &  $\alpha = 08^h 45^m 48^s,\delta = +45^{\circ} 01^{'} 17^{''}$ & $2.60\times 10^{-4}$ & 12 & $3.68\times 10^{-3}$ \newline 0.119 \newline\\ 
        1412 & \citet{Oort_and_van_Langevelde_1987} & WSRT & Hercules & 2 Centers & $4.12\times 10^{-3}$ & 13.73 & $7.03\times 10^{-4}$ \newline 0.637 \newline\\ 
        N/A & \citet{Oort_1987_PhD_Th} & N/A & N/A & N/A & N/A & N/A & N/A \newline\\ 
        1464.9 & \citet{Condon_1984} & VLA & Unnamed &  $\alpha = 08^h 52^m 15^s,\delta = +17^{\circ} 16^{'} 0^{''}$ & $3.67\times 10^{-3}$ & 19 & $7.95\times 10^{-3}$ \newline 0.133 \newline\\ 
        1490 & \citet{Mitchell_1985} & VLA & Unnamed, near NGP &  $\alpha = 13^h 0^m 37^s,\delta = +30^{\circ} 34^{'} 0^{''}$ & $3.67\times 10^{-3}$ & 17.5 & $6.80\times 10^{-3}$ \newline 0.134 \newline\\ 
        1465 & \citet{Coleman_1985} & VLA & Unnamed & $\alpha = 08^h 52^m 15^s,\delta = +17^{\circ} 16^{'} 0^{''} $& $3.67\times 10^{-3}$ & 5.8 & $2.77\times 10^{-5}$ \newline 4.1 \newline\\
        1411 & \citet{Condon_1982} & VLA & Unnamed &  $\alpha = 08^h 52^m 15^s,\delta = +17^{\circ} 16^{'} 0^{''}$ & $3.67\times 10^{-3}$ & 17.5 & 0.134 \newline 0.0124 \newline\\ \hline    
            \end{tabular}
            \label{tab:A.1}
\end{table*}

\begin{table*}
\renewcommand\thetable{A.1B} 
\caption{(cont'd) Summary and breakdown for the \citep{Windhorst_2003} collection. Coordinates given for papers before 2000 are as measured in J1950 and due to the age of some of the papers or their
status as PhD theses, some entries in the table are missing and are labeled as N/A.}
\begin{tabular}{p{1.0cm}p{3.0cm}p{1.5cm}p{2.0cm}p{3.0cm}p{1.5cm}p{1.5cm}p{2.0cm}} \\ \hline
Frequency  & Reference & Instrument & Field/Survey  & Field Location & Area & Beam FWHM & Confusion Source Density  \\ 
(MHz) & & & Name & & (Ster) & (Arcsec)  & Faintest Source Bin (mJy) \\ \hline
        5000 & \citet{Kuhr_1981} & Multiple & NRAO MPI Strong Source Surveys & Multiple Centers & 9.811 & 168 & $1.86\times 10^{-5}$ \newline 1064 \newline\\ 
        4850 & \citet{Maslowski_1984} & 100M MPIfR Telescope & Green Bank Region & Near $\alpha = 10^h 45^m 0^s,\delta = +47^{\circ} 21^{'} 15"$ & $9.18\times 10^{-3}$ & 168 & $6.26\times 10^{-3}$ \newline 11.4 \newline\\ 
        4755 & \citet{Owen_1983} & NRAO 300-ft Telescope & Unnamed & Strip centered at $\alpha = 12^h 30^m 0^s,\delta = +35^{\circ} 0^{'} 0^{''}$ & $6.91\times 10^{-2}$ & 173.5 & $6.76\times 10^{-4}$ \newline 83.3 \newline\\ 
        4755 & \citet{Ledden_1980} & NRAO 300-ft Telescope & Unnamed & Strip centered at $\alpha = 12^h 32^m 50^s,\delta = +35^{\circ} 0^{'} 0^{''}$ & $9.56\times 10^{-3}$ & 173.5 & $4.27\times 10^{-3}$ \newline 16.8 \newline\\ 
        4760 & \citet{Altschuler_1986} & NRAO 300-ft Telescope & Unnamed & Strip near $\delta = +33^{\circ}$ & $3.53\times 10^{-2}$ & 173.3 & $3.35\times 10^{-3}$ \newline 21.0 \newline\\ 
        5000 & \citet{Wrobel_1990} & VLA & Unnamed & Multiple Centers & $6.91\times 10^{-4}$ & 5 & $2.34\times 10^{-5}$ \newline 1.86 \newline\\ 
        4860 & \citet{Donnelly_1987} & VLA & Lynx 2 & Multiple Centers Near  $\alpha = 08^h 41^m 24^s,\delta = +44^{\circ} 42^{'} 37^{''}$ & 3.91E-05 & 16.5 & $1.70\times 10^{-3}$ \newline 0.197 \newline\\ 
        5000 & \citet{Fomalont_1991} & VLA & DEEPS2 &  $\alpha = 14^h 16^m 0^s,\delta = +52^ {\circ} 42^{'} 0" $ & $1.50\times 10^{-5}$ & 1.55 & $2.21\times 10^{-4}$ \newline 0.0176 \newline\\ 
        10700 & \citet{Seielstad_1983} & 40m Owens-Valley Radio Observatory & Unnamed & Multiple Centers & $1.58\times 10^{-3}$ & 180 & $6.80\times 10^{-3}$ \newline 76.4 \newline\\ 
        10000 & \citet{Aizu_1987} & 45m NRO Telescope - University of Tokyo & Unnamed & Large-Area Survey & 5.02 & 162 & $2.91\times 10^{-3}$ \newline 22.5 \newline\\ 
        8465 & \citet{Fomalont_2002} & VLA & SA13 and Hercules Field & 2 Centers & Area Dependent on Depth & 6 & $4.49\times 10^{-3}$ \newline 0.0137 \newline\\ \hline
            \end{tabular}
            \label{tab:Table_A.1B}
\end{table*}

\newpage

\subsection{Brief Discussion of Other Frequencies}
Here we discuss the data at the other six frequencies studied as part of this paper. At 150\,MHz, we display available data and the recently released data of \citet{Hardcastle_2021}. As a part of their survey, they compile source counts from 3 different fields sharing the same bin centers and flux range, so we averaged the corrected and uncorrected sets of data for each field to display them in one set, combining the errors in quadrature, and displaying the source counts spanning a larger range from the LoTSS counts on their own, while omitting data points that are clearly incomplete at the faint end. At 325\,MHz, a large error floor of 20\% is used to bring the surveys together within their error bars given the vertical offsets between the data. The data behaves well at 610\,MHz, 1.4\,GHz, and 5\,GHz, At 8.4\,GHz, a small amount of data and inconsistencies between the surveys make the structure of the radio EBL difficult to discern. 

\newpage

\onecolumn
\begin{figure}
\renewcommand{\thefigure}{A1}
\begin{multicols}{2}
\vspace*{0.000cm}

    \includegraphics[width=7.5cm, height = 9cm]{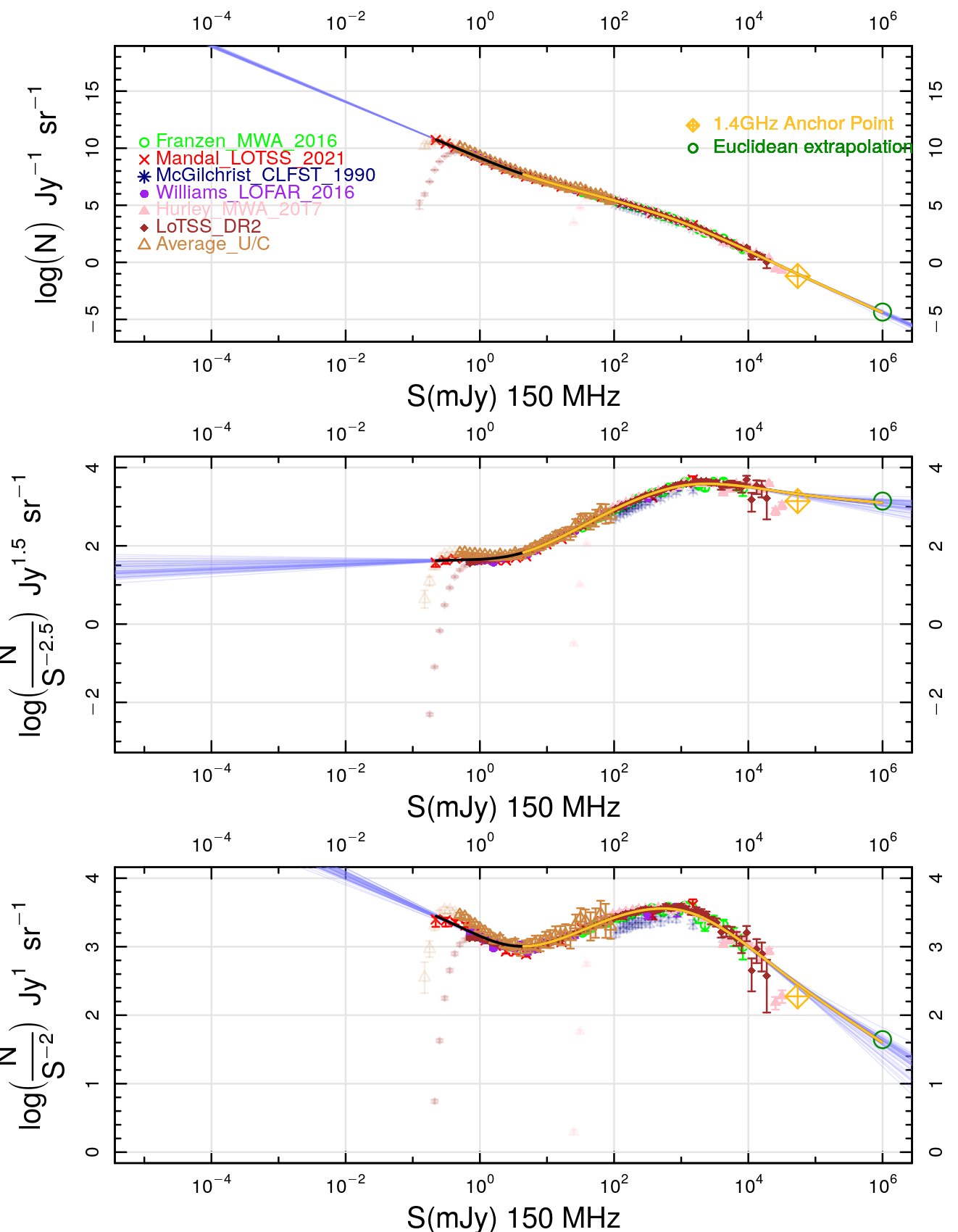}\par
    \includegraphics[width=7.5cm, height = 9cm]{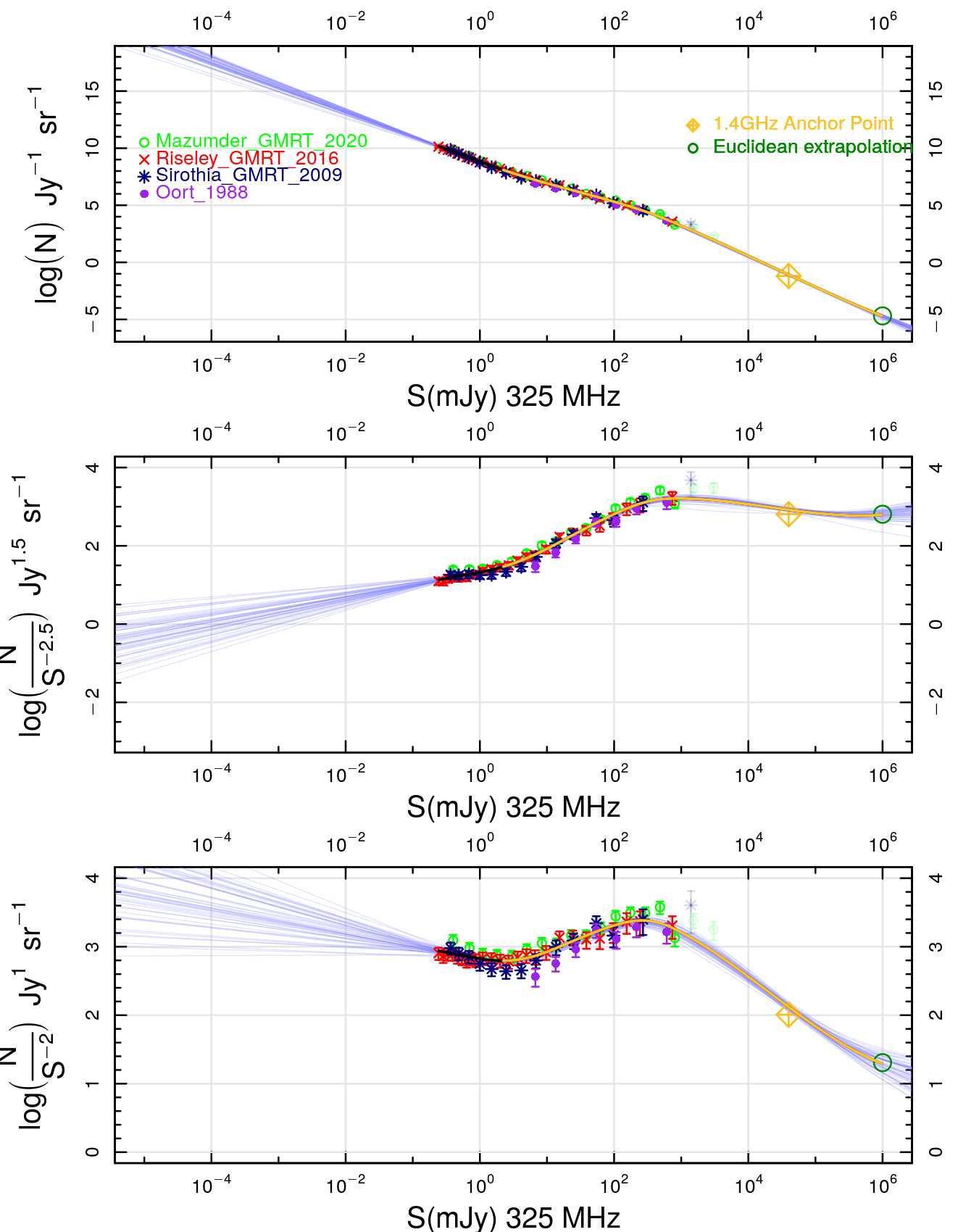}\par

\end{multicols}
\begin{multicols}{2}
    \includegraphics[width=7.5cm, height = 9cm]{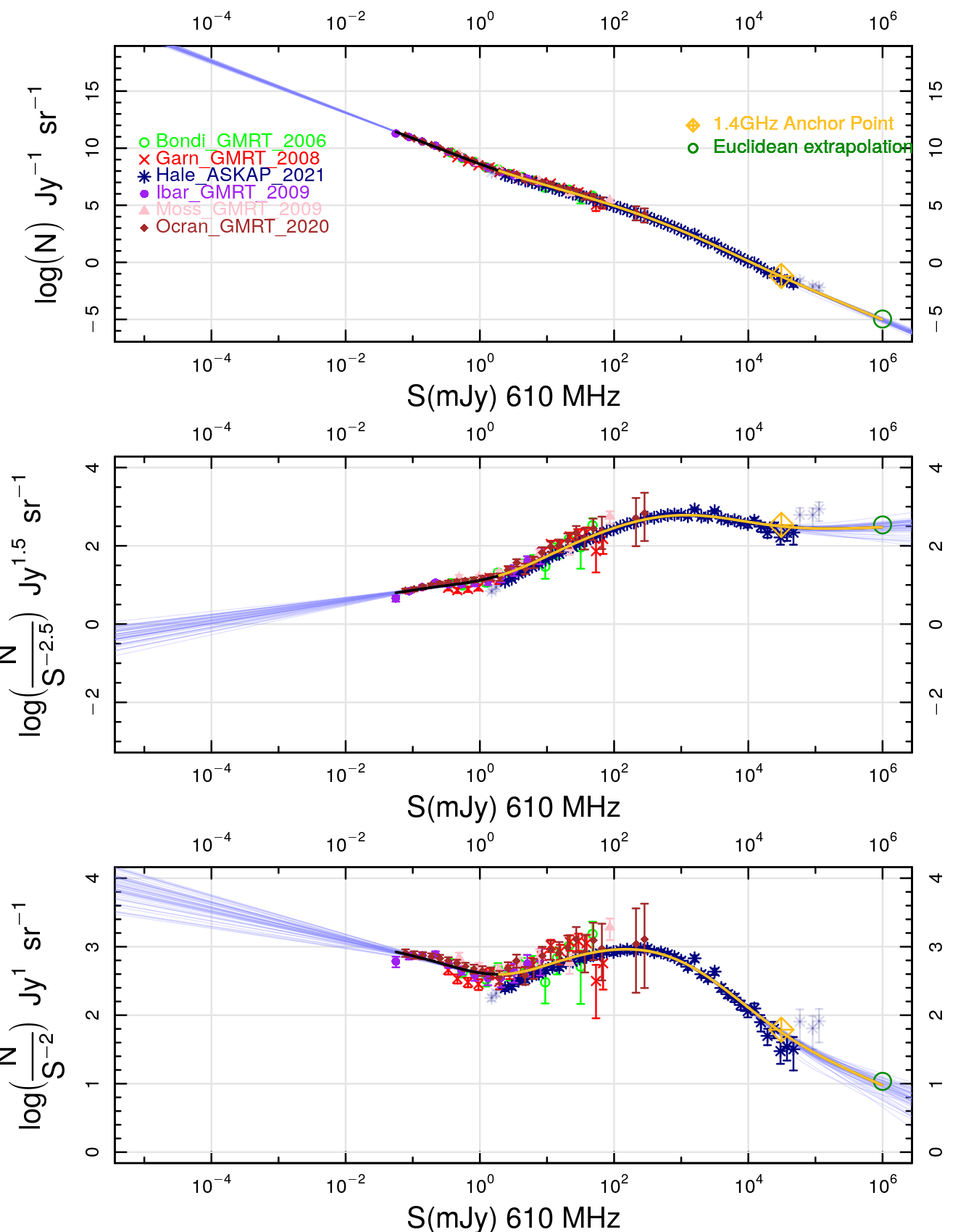}\par
    \includegraphics[width=7.5cm, height = 9cm]{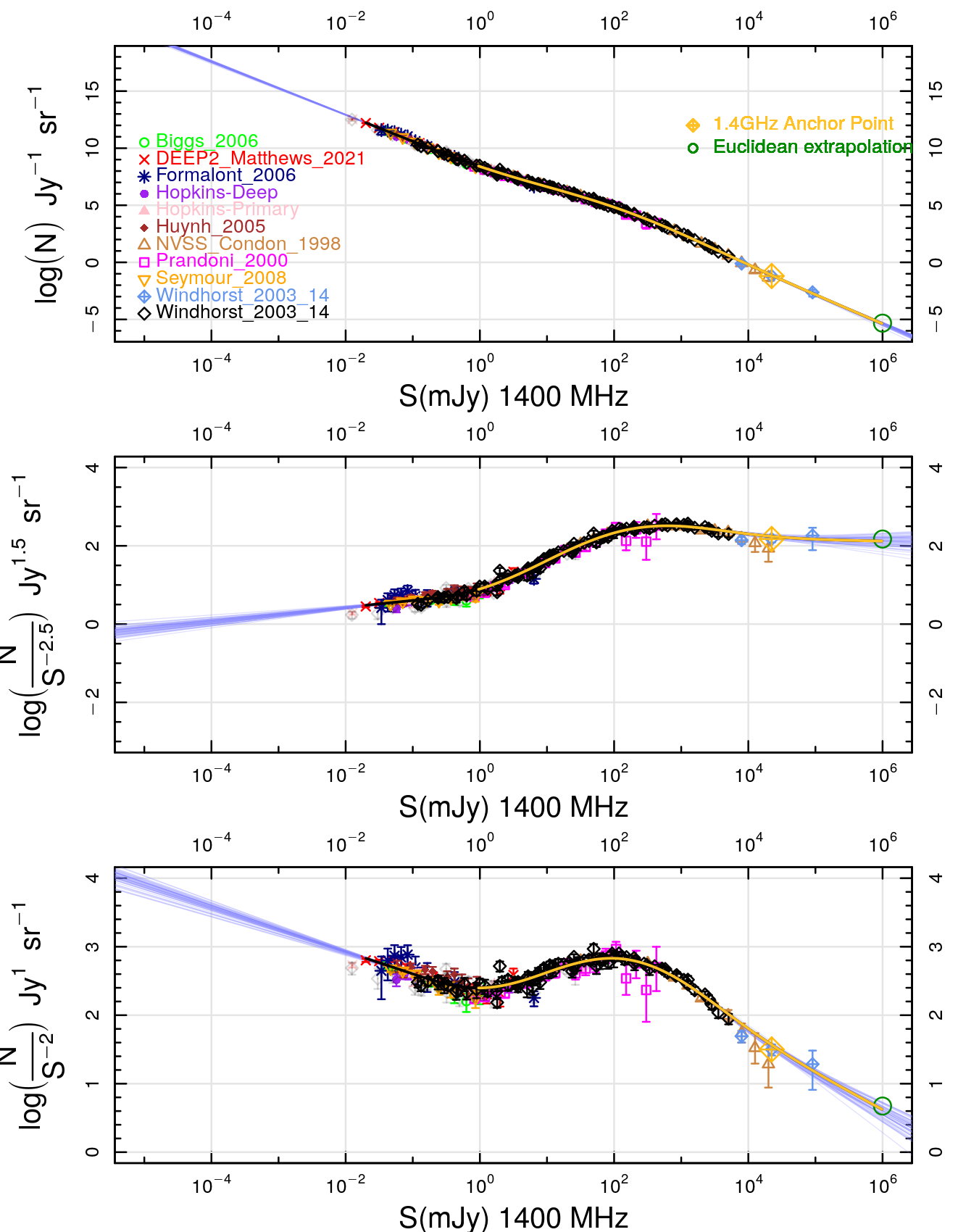}\par

\end{multicols}

\caption{Figures A1-A1-D show the same 3-panel display as Figure \ref{fig:fig2}. Convergence is not seen in the radio EBL in any other frequency, and thus only integration to the faintest data point is used to establish a lower limit.}
\label{fig:A1}
\end{figure}
\newpage
\clearpage

\newpage
\clearpage

\begin{figure}
\renewcommand{\thefigure}{A2}
\begin{multicols}{2}
\vspace*{0.000cm}

    \includegraphics[width=7.5cm, height = 9cm]{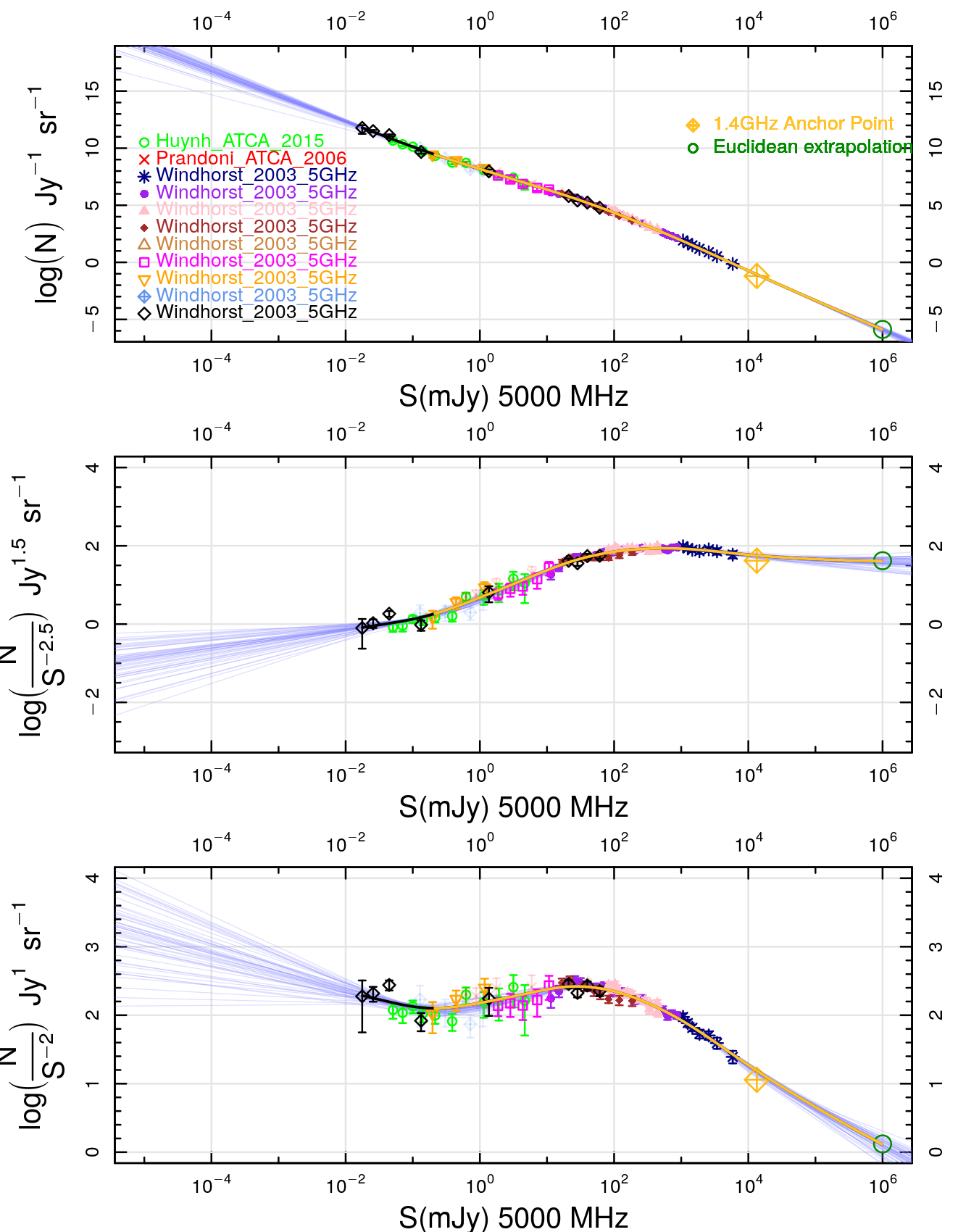}\par
    \includegraphics[width=7.5cm, height = 9cm]{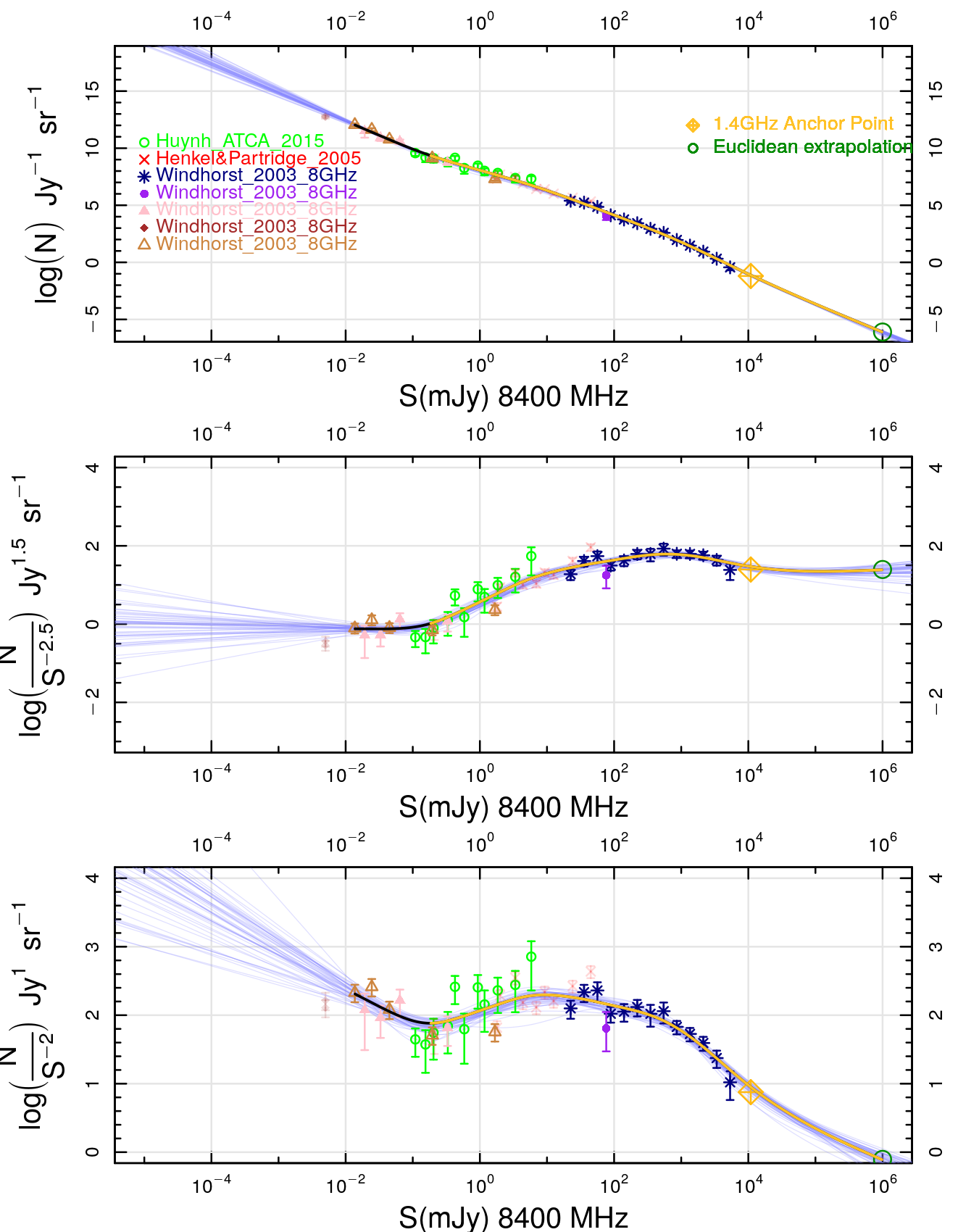}\par

\end{multicols}
\caption{Figures A2-A2-B show the same 3-panel display as Figure \ref{fig:fig2}. Convergence is not seen in the radio EBL in either frequency, and thus only integration to the faintest data point is used to establish a lower limit. A small amount of data at the higher frequencies results in larger errors as shown in \ref{tab:Table_2}.}
\label{fig:A2}
\end{figure}

\newpage
\clearpage

\begin{table*}
\renewcommand\thetable{A.2} 
\caption{The data used to derive the spectral index ($\alpha$) versus flux relationship at 610 MHz shown in Figure \ref{fig:fig1}b. Data and references are the same as used in \citet{Windhorst_2003}. The original $\alpha$ measurements are provided by \citet{Kapahi_1986}.}

\begin{tabular}{ccccc}
\hline
S 610 MHz (mJy) & Median $\alpha$ (610 - 1400 MHz) & \# Of Sources & Reference \\ \hline
17920 & $0.816^{+0.025}_{-0.025}$ & 77 & \citet{Robertson_1973} \\ 
8490 & $0.860^{+0.028}_{-0.028}$ & 83 &  \\
2560 & $0.866^{+0.014}_{-0.014}$ & 145 & \citet{Katgert-Merkelijn_1980}\\
2870 & $0.890^{+0.025}_{-0.025}$ & 72 &    \\
1810 & $0.895^{+0.024}_{-0.24}$ & 73 &     \\
2010 & $0.906^{+0.023}_{-0.023}$& 81 & \citet{Grueff_1979}   \\
1050 & $0.891^{+0.021}_{-0.021}$ & 115 &     \\
694 & $0.910^{+0.021}_{-0.021}$ & 124 &     \\
501 & $0.904^{+0.022}_{-0.022}$ & 94 & \citep{Kulkarni_1985A, Kulkarni_1985B}    \\
353 & $0.914^{+0.021}_{-0.021}$ & 90 &    \\
472 & $0.910^{+0.022}_{-0.022}$ & 105 & \citep{Davies_1973, Large_1981, Steppe_1984}   \\
231 & $0.890^{+0.025}_{-0.025}$ & 95 &     \\
167 & $0.908^{+0.025}_{-0.025}$ & 57 & \citep{Benn_1982, Benn_1984}    \\
39.4 & $0.802^{+0.033}_{-0.033}$ & 76 &     \\
14.1 & $0.787^{+0.038}_{-0.038}$& 55 &    \\
14.9 & $0.734^{+0.056}_{-0.056}$ & 32 & \citet{Oort_1988B}    \\
3.50 & $0.756^{+0.07}_{-0.07}$ & 20 & \citet{Donnelly_1987}    \\
0.803 & $0.65^{+0.10}_{-0.10}$ & 10 & \citet{Richards_1999}    \\
0.279 & $0.68^{+0.06}_{-0.06}$ & 25 &    \\
0.654 & $0.667^{+0.065}_{-0.065}$ & 22 & \citet{Fomalont_2006}   \\
0.170 & $0.929^{+0.056}_{-0.056}$ & 34 &    \\
0.0937 & $0.722^{+0.056}_{-0.056}$ & 33 &     \\
   \\ \hline
\end{tabular}
\label{tab:A.3}
\end{table*}

\begin{table*}
\renewcommand\thetable{A.3} 
\caption{The data used to derive the spectral index ($\alpha$) versus flux relationship at 4.86 GHz shown in Figure \ref{fig:fig1}. Data and references are the same as used in \citet{Windhorst_2003}.}

\begin{tabular}{cccccccc}
\hline
S 4.86 GHz (mJy) & Median $\alpha$ (1.4-5 GHz) & \# Of Sources  & S low (mJy) & S high (mJy) & Mean $\alpha$ & Reference \\ \hline
1817.0 & $0.41^{+0.06}_{-0.06}$ & 320 & 800 & 5000 & $0.53^{+0.03}_{-0.03}$ & \citet{Witzel_1979} \\ 
209.0 & $0.47^{+0.04}_{-0.04}$ & 120  & 67 & 800 & $0.49^{+0.05}_{-0.05}$ &  \citet{Pauliny-Toth_1978}\\
38.7 & $0.80^{+0.03}_{-0.03}$ & 212  & 15 & 100 & $0.25^{+0.03}_{-0.03}$  & \citet{Condon_1981}\\
2.57 & $0.68^{+0.11}_{-0.11}$ & 23 & 0.6 & 10.0 & $0.31^{+0.10}_{-0.10}$ & \citet{Fomalont_1984}    \\
0.687 & $0.42^{+0.10}_{-0.10}$ & 27 & 0.4 & 1.2 & $0.50^{+0.08}_{-0.08}$ & \citet{Donnelly_1987}    \\
0.197 & $0.40^{+0.09}_{-0.09}$& 27 & 0.1 & 0.4 & $0.57^{+0.08}_{-0.08}$ & \citet{Donnelly_1987}   \\
0.04 & $0.38^{+0.07}_{-0.07}$ & 29 & 0.016 & 0.100 & $0.60^{+0.07}_{-0.07}$ & \citet{Fomalont_1991}    \\
0.0462 & $0.35^{+0.15}_{-0.15}$ & 20 & 0.0175 & 0.121 & $0.50^{+0.20}_{-0.20}$ & \citet{Windhorst_1993}    \\
0.230 & $0.25^{+0.15}_{-0.15}$ & 12 & 0.0112 & 1.112 & $0.20^{+0.15}_{-0.15}$ & \citet{Richards_1999}    \\
0.037 & $0.50^{+0.12}_{-0.12}$ & 17 & 0.021 & 0.112 & $0.50^{+0.15}_{-0.15}$ &    \\
0.141 & $0.37^{+0.10}_{-0.10}$ & 19 & 0.00 & 0.00 & $0.28^{+0.08}_{-0.08}$ & \citep{Ciliegi_1999, Ciliegi_2003}   \\
1.817 & $0.81^{+0.14}_{-0.14}$ & 12 & 0.00 & 0.00 & $0.73^{+0.12}_{-0.12}$ &    \\
   \\ \hline
\end{tabular}
\label{tab:A.3B}
\end{table*}

\begin{table*}
\renewcommand\thetable{A.4} 
\caption{A sample of the compiled master data table used to generate all data shown in the 3-panel figures. Data is converted to the format shown in the top panel of Figure \ref{fig:fig2} and an upper and lower limit, along with the average of the upper and lower error is shown, all in the same format. Finally, individual points are flagged to classify if they are used in the calculations, and if not, why. A flag value of 0 indicates a good data point, 1 indicates a data point which is clearly incomplete or has one or fewer objects per bin, 2 indicates a point which display clear non-Euclidean behavior at the bright end, 3 indicates a survey not used for a particular reason, see Section \ref{Section_2.2}, and 4 indicates a reference which could not be located as part of the Windhorst compendium \citet{Windhorst_2003}. Flux values shown are corrected from their original frequency as described in Section 2.3.}

\begin{tabular}{cccccccccc}
\hline
Unique ID & Frequency (MHz) & Survey Name & S(mJy) & log(N) & avg error log(N) & up lim log(N) & low lim log(N) & Good Point \\ \hline
1 & 150 & Franzen\_2016 & 7939.7 & 1.2104 & 0.17945 & 1.3852 & 1.0262 & 0 \\ 
1 & 150 & Franzen\_2016 & 5786.2 & 1.6294 & 0.13096 & 1.7584 & 1.4964 & 0 \\
... & ... & ... & ... & ... & ... & ... & ... & 0 \\
11 & 610 & Hale\_ASKAP\_2021 & 149506 & -2.6468 & 0.2385 & -2.4707 & -2.9479 & 2 \\
   \\ \hline
\end{tabular}
\label{tab:A.4}
\end{table*}

\end{document}